\documentclass[%
 aip,
 amsmath,amssymb,
 reprint,%
]{revtex4-2}

\usepackage{graphicx}% Include figure files
\usepackage{dcolumn}% Align table columns on decimal point
\usepackage{bm}% bold math
\usepackage[utf8]{inputenc}
\usepackage[T1]{fontenc}
\usepackage{mathptmx}
\usepackage{etoolbox}
\usepackage{ulem}

\RequirePackage{kvdefinekeys, kvsetkeys, kvoptions}
\RequirePackage{iftex}
\RequirePackage{ifthen}
\usepackage{float}
\usepackage{color}
\usepackage{makecell}
\usepackage{ulem}

\makeatletter
\def\@email#1#2{%
 \endgroup
 \patchcmd{\titleblock@produce}
  {\frontmatter@RRAPformat}
  {\frontmatter@RRAPformat{\produce@RRAP{*#1\href{mailto:#2}{#2}}}\frontmatter@RRAPformat}
  {}{}
}%
\makeatother
\makeatletter
\AtBeginDocument{\@ifpackageloaded{natbib}{\ifNAT@numbers\if@filesw\immediate\write\@auxout{\string\global\string\NAT@numberstrue}\fi\fi}{}}
\makeatother

\begin{document}

\preprint{AIP/123-QED}

\title{ Efficiency and Forward Voltage of  Blue and Green Lateral LEDs with V-shape defects and Random Alloy Fluctuation in Quantum Wells}% Force line breaks with \\

\author{Cheng-Han Ho}

\affiliation{Graduate Institute of Photonics and Optoelectronics, National Taiwan University, Taipei 10617, Taiwan}

\author{James S. Speck}

\affiliation{Materials Department, University of California,
Santa Barbara, California 93106, USA
}

\author{Claude Weisbuch}

\affiliation{Materials Department, University of California,
Santa Barbara, California 93106, USA
}

\affiliation{Laboratoire de Physique de la Mati\`ere Condens\'ee, Ecole Polytechnique, 
CNRS, IP Paris, 91128 Palaiseau Cedex, France
}

\author{Yuh-Renn Wu}%
 \altaffiliation{Author to whom correspondence should be addressed: yrwu@ntu.edu.tw%\\This line break forced with \textbackslash\textbackslash
}%
\affiliation{Graduate Institute of Photonics and Optoelectronics, National Taiwan University, Taipei 10617, Taiwan}
\affiliation{Electronic and Optoelectronic System Research Laboratories, Industrial Technology Research Laboratories, Hsinchu, Taiwan. }

\date{\today}% It is always \today, today,

\begin{abstract}
For nitride-based blue and green light-emitting diodes (LEDs), the forward voltage  $V_\text{for}$ is larger than expected, especially for green LEDs. This is mainly due to the large barriers to vertical carrier transport caused by the total polarization discontinuity at multiple quantum well  and quantum barrier interfaces. The natural random alloy fluctuation in  QWs has proven to be an important factor reducing $V_\text{for}$. However, this does not suffice in the case of green LEDs because of their larger polarization-induced barrier.  V-defects have been proposed as another key factor in reducing  $V_\text{for}$ to allow laterally injection into multiple quantum wells (MQWs), thus bypassing the multiple energy barriers incurred by vertical transport. In this paper, to model  carrier transport in the whole LED, we consider  both random-alloy and V-defect effects. A fully two-dimensional drift-diffusion charge-control solver is used to model both effects. The results indicate that the turn-on voltages for blue and green LEDs are both affected by random alloy fluctuations and V-defect density. For  green LEDs, $V_\text{for}$  decreases more due to  V-defects, where the smaller polarization barrier at the V-defect sidewall is the major path for lateral carrier injection. Finally, we discuss how  V-defect density and size  affects the results. 
\end{abstract}

\maketitle

%\tableofcontents

\section{Introduction}

Blue nitride-based  light-emitting diodes (LEDs) with phosphor have become the major  white-light source. However,  high-efficiency or micro-LED display applications require green- and red-light sources and, for ultimate efficiency, white-light lamps based on RGB color mixing. Currently, the peak internal quantum efficiency ($IQE$) of blue LEDs is over 90$\%$ at a current density of 1--10 A/cm$^2$.\cite{weisbuch2019search} However, the $IQE$ of green LEDs remains low, which is likely due to the strong quantum-confined Stark effect induced by the polarization charge and the large defect density\cite{weisbuch2019search,PhysRevLett.116.027401,6935331}. In addition, the forward voltage $V_\text{for}$ of green LEDs  also exceeds the expected 2.3~V, \cite{lynsky2020barriers} which also limits the wall plug efficiency. 

III-nitrides in the common (0001) orientation have polarization-induced barriers at the InGaN/GaN quantum well/quantum barrier (QW/QB) interface due to  spontaneous and piezoelectric polarization differences between the well and barrier materials.\cite{mukund2014ingan} The polarization discontinuity at the QW/QB interface results in significant electric fields in the QWs and QBs---often referred to as ``piezofields.'' For green LEDs, the polarization discontinuity at the QW/QB interface increases due to the increased lattice mismatch between the  InGaN and GaN layers. 

The influence of polarization-induced barriers has been discussed in many studies\cite{lheureux20203d,qwah2020theoretical,lynsky2020barriers} where carriers are easily blocked by the extra potential barrier. Our previous studies of blue LEDs \cite{yang2014influence,wu2012analyzing} show that fluctuations in indium composition   provide extra paths for  carrier injection. Including random indium composition fluctuations in models thus helps to explain the lower $V_\text{for}$ in blue LEDs compared with models without fluctuations. \cite{chen2018three,der2015influence} 

To further reduce $V_\text{for}$ in situations with larger barriers, such as in green LEDs, it is important to  either reduce the polarization field or find an alternative way to inject  carriers. The most significant method is by controlling the V-shaped defect (V-defect) opened at the InGaN/GaN MQW active regions to provide an alternative way to inject carriers. Most V-defects are the result of dislocation lines that  form under certain growth conditions. \cite{lester1995high,voronenkov2013nature,yoshida2015impact} Unlike the threading dislocation (TD) centers, which are typical nonradiative centers, the inclined sidewalls of  V-defects  reportedly  assist  carrier injection into QWs. \cite{li20163d,cho2013quantum,quan2016effect} Since both random alloy fluctuations and V-defects provide concurrent paths for carrier injection, it is important to model both effects in carrier injection simultaneously. The importance of considering both mechanisms is obvious as the paper leads to the conclusion that random alloy fluctuations suffice to lead to efficient low-voltage injection in blue LEDs while V-defects are needed to achieve this in longer wavelength LEDs in the green spectral range.

Given that V-defects play an important role in carrier injection, it is important to understand  the role of random alloy fluctuations and of V-defects in carrier injection. The V-defect and random alloy effects may be simulated separately in three-dimensional simulations. \cite{yang2014influence,li20163d} On the one hand,  V-defects range from a few hundred nanometers in size to the $\mu$m scale, allowing for large mesh sizes. On the other hand, random alloy fluctuations occur on a scale of a few nanometers, requiring  mesh elements to be as small as possible to account for such small-scale fluctuations. Including both effects in a three-dimensional simulation requires unreasonable computer memory  (> a few TB) and computing times. However, to understand the influence of random alloy fluctuation and V-defect on the efficiency of LEDs, especially for green LEDs, it is essential to include both effects simultaneously. Recent results on the lateral carrier diffusion\cite{shen2021three} in QW show that the random alloy fluctuations strongly limits carrier diffusion. Hence, if the carriers are only injected through V-defects, the current spreading to the whole QW area from the V-defect will be limited, which is observed without considering random alloy fluctuation. Furthermore, the improvement in turn-on voltage induced by V-defects, critical to the wallplug efficiency, can be jeopardized by the adverse action of the threading dislocation within the V-defects as nonradiative centers on the $IQE$. The balance between the reduction of $V_\text{for}$ by increasing the V-defect density and the simultaneous decrease in $IQE$ must be explored. A compromise to the present difficulties of 3D simulations is to work on two-dimensional (2D) simulations that include random alloy fluctuations and V-defects. Some issues might be missed in the 2D model, such as the estimation of carrier quantum confinement effects occurring in 3D. This may affect the estimation of carrier overlap in random alloy fluctuating potentials. The V-defect filling factor of 2D and 3D models may not be equal and needs some translation. However, the 2D model enables us to calculate the whole LED structure, including the current crowding effect. This helps us to understand the simultaneous influences of V-defect density and random alloy fluctuation to $V_\text{for}$ or and $IQE$. In this paper, we  discuss the influence of  V-defects and random alloy fluctuations with such an alternative 2D model. 

V-defects may be controlled by tuning the growth conditions. Often, V-defects are formed in the widely used InGaN/GaN short periodic superlattice that is grown before growing active InGaN/GaN QWs. \cite{zhou2017effect} The general V-defect density is about 1$\times$10$^7$ cm$^{-2}\sim 5\times10^8$~cm$^{-2}$, and the typical diameter is about 30--250 nm for five QWs in blue LEDs. \cite{lester1995high,voronenkov2013nature,yoshida2015impact, zhou2017effect} Hence, in this paper, we will cover the V-defect density from 0 to $\sim$ 6.25$\times10^8$ cm$^{-2}$. Furthermore, V-defects form an inverted hexagonal pyramid with six inclined $\left\{10\bar{1}1\right\}$ semipolar sidewalls. \cite{wu1998structural} Studies show that the indium composition in $c$-plane QWs is greater than in the inclined QWs in the V-defect region. \cite{hu2012effect} Therefore, the bandgap of QWs in the inclined sidewall of V-defects is larger than the bandgap of QWs in the $c$ plane. The polarization discontinuity between the semipolar $\left\{10\bar{1}1\right\}$ InGaN sidewall QW and QB is much smaller than the $c$-plane QW and is in the opposite direction. \cite{romanov2006strain} Thus, holes can be injected into the QWs in the $c$ plane through the sidewall V-defect QWs with lower barrier potentials. \cite{jiang2019efficient} The additional paths provided decreased $V_\text{for}$  and increase the carrier-injection efficiency. \cite{lynsky2021role,han2013improvement,ren2016analysis} However, V-defects form at TDs, and a high density of nonradiative centers at the center of V-defects could be a drawback. Thus, the  density and diameter of the V-defects in the LED should be optimized to minimize this counter-acting effect. \cite{bouveyron2019v}

In this paper, we  use simulations to investigate how random fluctuations and  V-defects affect InGaN-based LEDs. As  mentioned above,  computer-memory limitations oblige us to use  2D simulations with fluctuations instead of three-dimensional simulations. In 2D simulations, the size of the simulated structure may be increased to the micron scale to approach the real device structure of LEDs. In the following, we discuss the 2D structure with different V-defect densities and diameters (lateral size). The forward voltage $V_\text{for}$ and efficiencies (electrical and $IQE$) are discussed and explained by the interplay between lateral carrier injection through the V-defect sidewalls and subsequent carrier diffusion within the planar QW regions.

\section{Methodology}

To simulate random alloy fluctuations in devices, we  use an in-house-developed 2D drift-diffusion charge control solver  to determine the electrical carrier distribution, recombination rates, and other properties. First, to account for the random alloy fluctuations and V defects, we construct a mesh for a device with a V-defect. The mesh size at the QW region is 0.5 nm in the $x$ direction and 0.1 nm in the $z$ direction. The mesh size is larger in the $p$ and $n$ layers to save computer memory. We then use a random number generator to create the random atom distribution in the alloy regions \cite{di2020simulating}  and subsequently use  Gaussian averaging  to assign the local indium composition in the QWs. \cite{filoche2017localization, PhysRevB.95.144205, li2017localization}

Figure \ref{part2 flow chart} shows the indium composition fluctuations in the QWs, where the bandgap  depends on the indium composition. Note that, in the 2D simulation, we cannot calculate the real strain distribution with a strain solver because we have  neglected  the third dimension. Thus, fully compressive strain is assumed when calculating  the lattice size of the GaN buffer layer based on  the local indium composition. We then calculate the local material parameters such as bandgap,  conduction- and valence-band energies, polarization according to the local indium composition, and strain state. In addition, we use the localization landscape (LL) model \cite{filoche2017localization, PhysRevLett.116.056602,filoche2012universal} shown in Eq. (\ref{eq2})  instead of the Schrödinger equation solver  to obtain the effective quantum potential seen by the carriers. \cite{li2017localization}

\subsection{Methodology of analyzing two-dimensional solver}
To simulate the 2D Poisson, drift-diffusion, and LL equations, the following equations were used:\\
\begin{eqnarray}
  \bm\nabla\cdot\left(\varepsilon\bm\nabla\varphi\right)=q\left(n-p+N_\text{A}^--N_\text{D}^+\pm\rho_{\text{pol}}\right), 
  \label{eq1}\\
  \left[-\frac{\hbar^2}{2}\bm\nabla\left(\frac{1}{m_{\text{e,h}}^\ast}\bm\nabla\right)+E_{\text{c,v}}\right]u_{\text{e,h}}=1,
  \label{eq2}
\end{eqnarray}
\begin{eqnarray}   
  n&=&\int_{1/u_\text{e}}^{+\infty}{\text{LDOS}\left(E\right)\frac{1}{1+\exp\left(\frac{E-E_{\text{F}n}}{k_\text{B}T}\right)}dE},
  \label{eq3} \\
  p&=&\int_{1/u_\text{h}}^{-\infty}{\text{LDOS}\left(E\right)\frac{1}{1+\exp\left(\frac{E_{\text{F}p}-E}{k_\text{B}T}\right)}dE},
  \label{eq4}
\end{eqnarray}
\begin{eqnarray}   
  J_{n,p}&=&\mu_{n,p}\left(n,p\right)\bm\nabla E_{\text{F}_{n,p}},
  \label{eq5}\\
  \frac{1}{q}\bm\nabla\left(J_{n,p}\right)&=& R_{n,p}-G_{n,p},
  \label{eq6}
\end{eqnarray}
\begin{eqnarray}    
   R&=&R_\text{SRH}+B_0np+C_0\left(n^2p+np^2\right),
  \label{eq7}\\
  R_\text{SRH}&=&\frac{np-n_i^2}{\tau_{n0}\left(p+n_i\right)+\tau_{p0}\left(n+n_i\right)}.
  \label{eq8}
\end{eqnarray}

In these equations, $\varphi$ is the electrostatic potential in the structure,  $n$ and $p$ are the free electron and hole concentrations, and $N_\text{A}^-$ and $N_\text{D}^+$ are the ionized acceptor and donor concentrations determined by the ionization energy (and position in the junction and junction electric field), respectively. $\rho_{\text{pol}}$ is the polarization charge that has been computed from the divergence of the total polarization, and $\frac{1}{u_\text{e}}$ and $\frac{1}{u_\text{h}}$ are the effective quantum potentials of electrons and holes calculated by the LL equations, \cite{filoche2012universal, filoche2017localization} respectively. 

First, we calculate $\varphi$ by using the Poisson equation (\ref{eq1}) and obtain the local band extrema $E_\text{c}$ and $E_v$ through  the local $\varphi$ and alloy composition. Then, the LL equations  (\ref{eq2}) are solved to yield $
{u_\text{e}}$ and $
{u_\text{h}}$ and  the effective quantum potentials  $\frac{1}{u_\text{e}}$ and $\frac{1}{u_\text{h}}$. The carrier distributions are calculated by using Eqs. (\ref{eq3}) and  (\ref{eq4}) , the effective quantum potentials $\frac{1}{u_\text{e}}$ and $\frac{1}{u_\text{h}}$, and the quasi-Fermi levels $E_{\text{F}n}$ and $E_{\text{F}p}$. Finally, we determine $E_{\text{F}n}$ and $E_{\text{F}p}$ from Eqs. (\ref{eq5}) and  (\ref{eq6}). The recombination rate is given by Eq. (\ref{eq7}), and includes radiative recombination ($IQE$), Auger recombination, and Shockley–Read–Hall (SRH) recombination. $B_0$ is the radiative recombination rate. Equation (\ref{eq8}) is the SRH recombination equation, where $\tau_n$ and $\tau_p$ are the nonradiative carrier lifetimes of electrons and holes, respectively. $C_0$ is the Auger recombination rate coefficient, which is the recombination of three particles, either directly or mediated by phonons. \cite{kioupakis2011indirect}

\begin{figure}[tb]
\centerline{\scalebox{.35}{\includegraphics{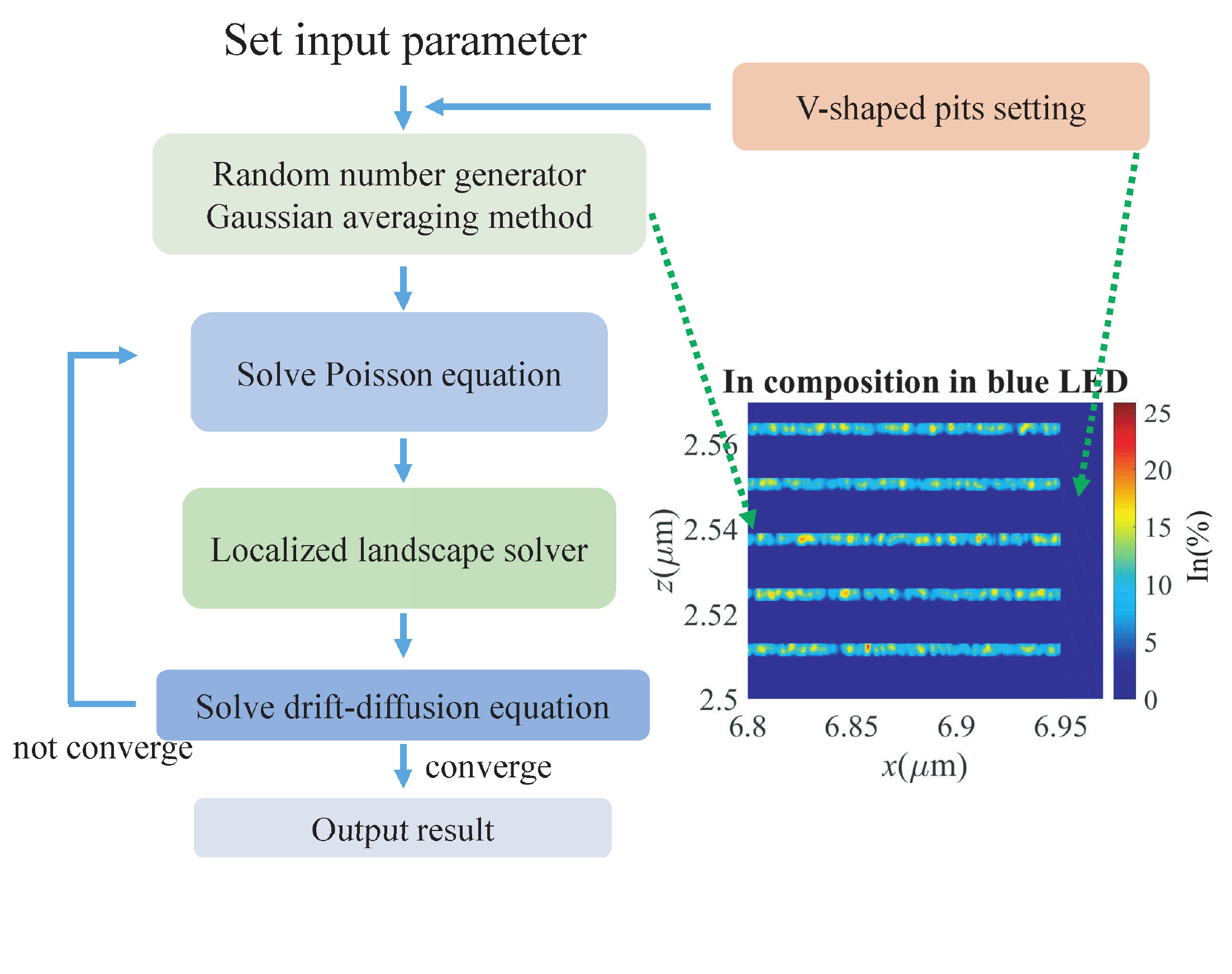}}}
\caption{Flow chart of simulation model. The mesh with V-defect shapes is put in the simulation program. The random alloy map is generated by using a random number generator and then put in the solver to realize convergence after the required number of feedback loops. The right panel shows the indium composition map in the MQWs.}
\label{part2 flow chart}
\end{figure}

Next, we construct the mesh according to the position in the structure; each mesh point has different parameters. To simulate V-defects in the structure, the mesh is readjusted according to the V-defect diameter and density. Finally, these equations are solved self-consistently. Figure \ref{part2 flow chart} shows a detailed flowchart of the simulation.

\subsection{Device structures and parameters}

\begin{figure}[tb]
\centerline{\scalebox{0.17}{\includegraphics{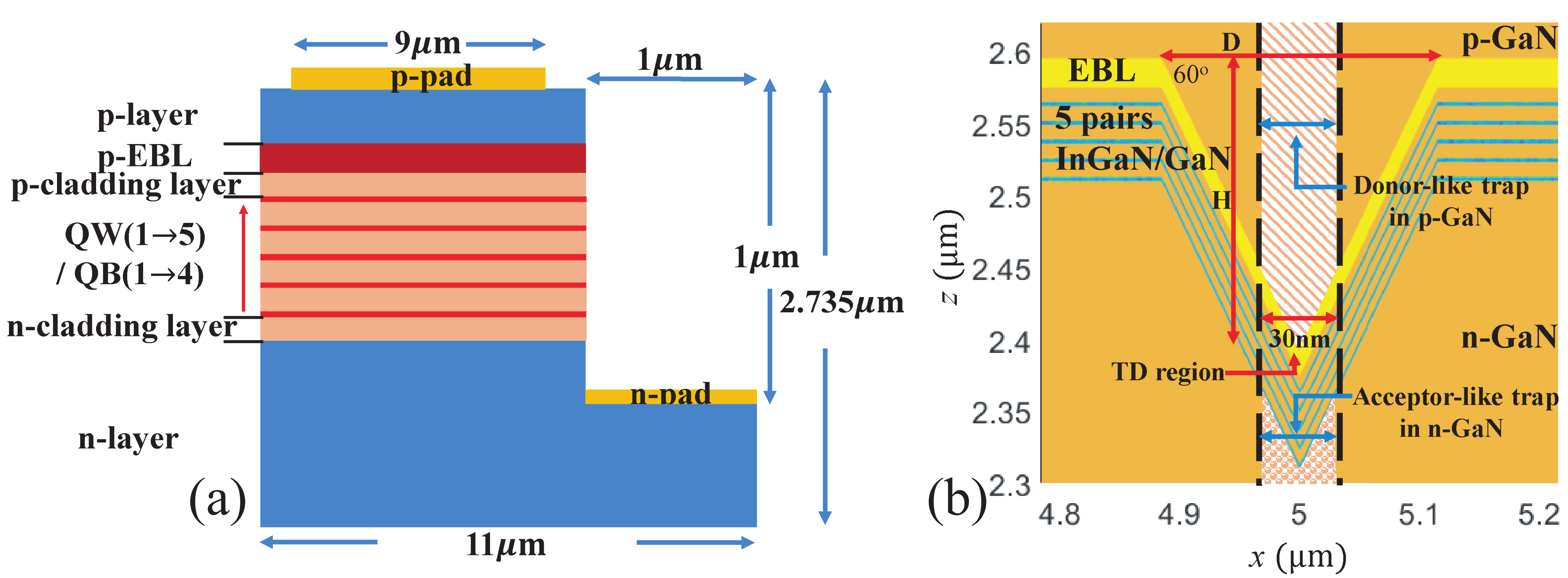}}}
\caption{(a) Simulated structure of  lateral LED. (b) Definition of parameters for  V-defect structure.}
\label{lateral LED and V-pit definition}
\end{figure}

To discuss the V-defect, we  analyze the blue and green LEDs with the structure shown in Fig. \ref{lateral LED and V-pit definition}(a). The chip lateral size is 11 $\mu$m and the region of multiple quantum wells (MQWs) is 10 $\mu$m wide, which is also defined as the total region. Since our mesh element size is quite small compared to the size of the device, we use about $1.1\times10^7$ nodes  to simulate the 11 $\mu$m chip, which requires 55 GB of memory to run the 2D simulation. The $p$ contact (in common $p$-side LEDs, this layer is made from indium tin oxide) is about 9 $\mu$m wide at the center of the active region, and the $n$ pad is about 1 $\mu$m wide. 

For the blue and green LEDs, we consider a MQW LED with a 0.14 $\mu$m top $p$-GaN layer doped at $3\times10^{19}$~cm$^{-3}$, a $p$-AlGaN electron blocking layer (EBL) and $p$-GaN cladding layer both doped at $2\times10^{19}$~cm$^{-3}$, five QWs and  four QBs between the QWs are unintentionally doped  $n$ type at $1\times 10 ^{17}$cm$^{-3}$, one 10 nm $n$-cladding layer and then a 2.5 $\mu$m $n$-GaN layer doped at $5\times 10 ^{18}$~cm$^{-3}$. The doping density the of $n$-type cladding layer is $1.0\times 10 ^{19}$~cm$^{-3}$. For the green LED, the first two QBs are $n$ doped at $1.0\times 10 ^{18}$~cm$^{-3}$ to partially screen the piezoelectric field for better vertical carrier injection. We also ran a simulated case without doping in the first two QBs (and with no V-defects) for comparison. In addition,  $\tau_{n}$ and $\tau_{p}$ for the blue and green LEDs are 100  and 10 ns, respectively, because the green LED has a higher In content in the InGaN than  the blue LED and thus a lower growth temperature, which leads to a higher density of SRH centers in the real device. The detailed parameters of the structures are shown in Table \ref{LED_table}. For the blue LED, the average indium composition in the QWs is 13$\%$ (In$_{0.13}$Ga$_{0.87}$N). For the green LED, the average indium composition in the QWs is 22$\%$ (In$_{0.22}$Ga$_{0.78}$N). Additionally, alloy fluctuations in the InGaN layers are included in both the blue and green LEDs.
The $B_{0}$ is $2$ $\times$ $10^{-11}$ cm$^{3}$/Vs and the $C_{0}$ is $2$ $\times$ $10^{-31}$ cm$^{6}$/Vs. These two parameters are not changed in the simulation since we focus on V-defect and random alloy fluctuation influences. The influence of overlap due to the different QCSE and disorder is included in the term of $n(r)\times p(r)$.

\begin{table*}
\caption{Basic parameters of blue and green LEDs.}
\begin{ruledtabular}
\begin{tabular}{cccccccc}
 Area & Material       & Thickness &  \makecell[c]{Doping\\blue ,  green}& $E_\text{a}$ & \makecell[c]{e, h\\mobility}   &  \makecell[c]{$\tau_{n}$, $\tau_{p}$\\blue }  &  \makecell[c]{$\tau_{n}$, $\tau_{p}$\\  green}\\ 
&unit    & nm & 10$^{18}$cm$^{-3}$ & meV  & cm$^2$/V\,s &ns  &ns \\ \hline
$p$ layer & $p$-GaN   &140 & 30, 30 & 180&100, 5&100& 10\\    
$p$ EBL & $p$-AlGaN & 20 & 20, 20 & 264&100, 5 &100& 10\\    
$p$ cladding layer & $p$-GaN     & 10 & 20, 20  & 180&100, 5 &100& 10\\    
QW(3,4,5) & $n$-InGaN   & 3  & 0.1, 0.1 & 25&150, 10 &100& 10\\    
QB(3,4) & $n$-GaN     & 10 & 0.1, 0.1 & 25&350, 10 &100& 10\\
QW(1,2) & $n$-InGaN   & 3  & 0.1, 0.1 & 25&150, 10 &100& 10\\    
QB(1,2) & $n$-GaN     & 10 & 0.1, 1 & 25&350, 10 &100& 10\\   
$n$ cladding layer & $n$-GaN     & 10 & 10, 10 & 25 & 200, 10 &100& 10\\    
$n$ layer & $n$-GaN     & 2500 & 5, 5 & 25 & 200, 10 &100& 10\\ 
\end{tabular}
\end{ruledtabular}
\label{LED_table}
\end{table*}

Figure 2(b) shows the geometrical definition of V-defects, where H and D are the depth and diameter of the V-defect, respectively. We  use these parameters in the following discussion. The region between the vertical black dotted lines correspond to the TD along the direction  $\left\{0001\right\}$ with a 30 nm width at the center of the V-defect, and the V-defect sidewall QWs are inclined about 60$^\circ$  according to both transmission electron microscopy (TEM) and scanning  TEM,\cite{le2012carriers,wu2014electroluminescence} which is consistent with early work showing that the sidewalls are $\left\{10\bar{1}1\right\}$ planes. In addition, the thickness of the $c$-plane QWs is 3 nm. The inclined sidewall QW is also 3 nm thick in the z direction, which corresponds to a thickness of 1.73 nm ($3\text{ nm}\times\cos60^\circ $)  along the normal  to the inclined plane. \cite{shiojiri2006structure} The indium composition in the inclined QWs depends on the composition in $c$-plane QWs. For example, for blue LEDs, the indium composition of inclined QWs is 8$\%$. \cite{chang2015manipulation,tomiya2011atomic} For  green LEDs, the indium composition in the inclined QWs is about 16$\%$. \cite{zhou2018effect} In addition, the TDs are defects induced by the lattice and chemical mismatch between the sapphire substrate and GaN. \cite{kumakura2005minority,karpov2002dislocation} Thus, we include TD-associated trap states in the TDs regions, which  trap either electron or holes and are shown as the cross-hatched areas in Fig. \ref{lateral LED and V-pit definition}(b). The density of TD-associated trap states is about $1\times 10 ^{18}$ cm$^{-3}$ corresponding to ~one trap state per $c$ translation along the dislocation line within the 30 nm diameter TD ranges.\cite{Robertson_2019,Qwah_2021} The trap energy level ($E_\text{t}$) is  1.14 eV below $E_\text{c}$ for donor-like traps and 2.5 eV below $E_\text{c}$ for acceptor-like traps, as shown in Table \ref{lateral_LED_trap}. The trap lifetimes are also given in Table \ref{lateral_LED_trap}. \cite{Qwah_2021}

\begin{table}
\caption{Trap parameters in the threading dislocation (TD) center.}
\begin{ruledtabular}
\begin{tabular}{cccccc}
Type  \\ \hline
Donor-like \\
&Trap density  (1/cm$^3$)  &  $1\times 10 ^{18}$   \\            
&Trap $\tau_{n}$,$\tau_{p} $  (ns) &  1.01  \\
&$E_\text{t}$ below $E_\text{c}$ (eV) &  1.14\\\hline
Acceptor-like \\
&Trap density (1/cm$^3$) &  $1\times$10$^{18} $ \\
&Trap $\tau_{n}$,$\tau_{p}$ (ns)  &  0.38  \\
&$E_\text{t}$ below $E_\text{c}$ (eV) &  2.5 \\
\end{tabular}
\end{ruledtabular}
\label{lateral_LED_trap}
\end{table}

\section{Results and discussion}

\subsection{Influence of random alloy fluctuations, V-defects, and threading dislocations for the current  path}

\begin{figure}[tb]
\centerline{\scalebox{0.15}{\includegraphics{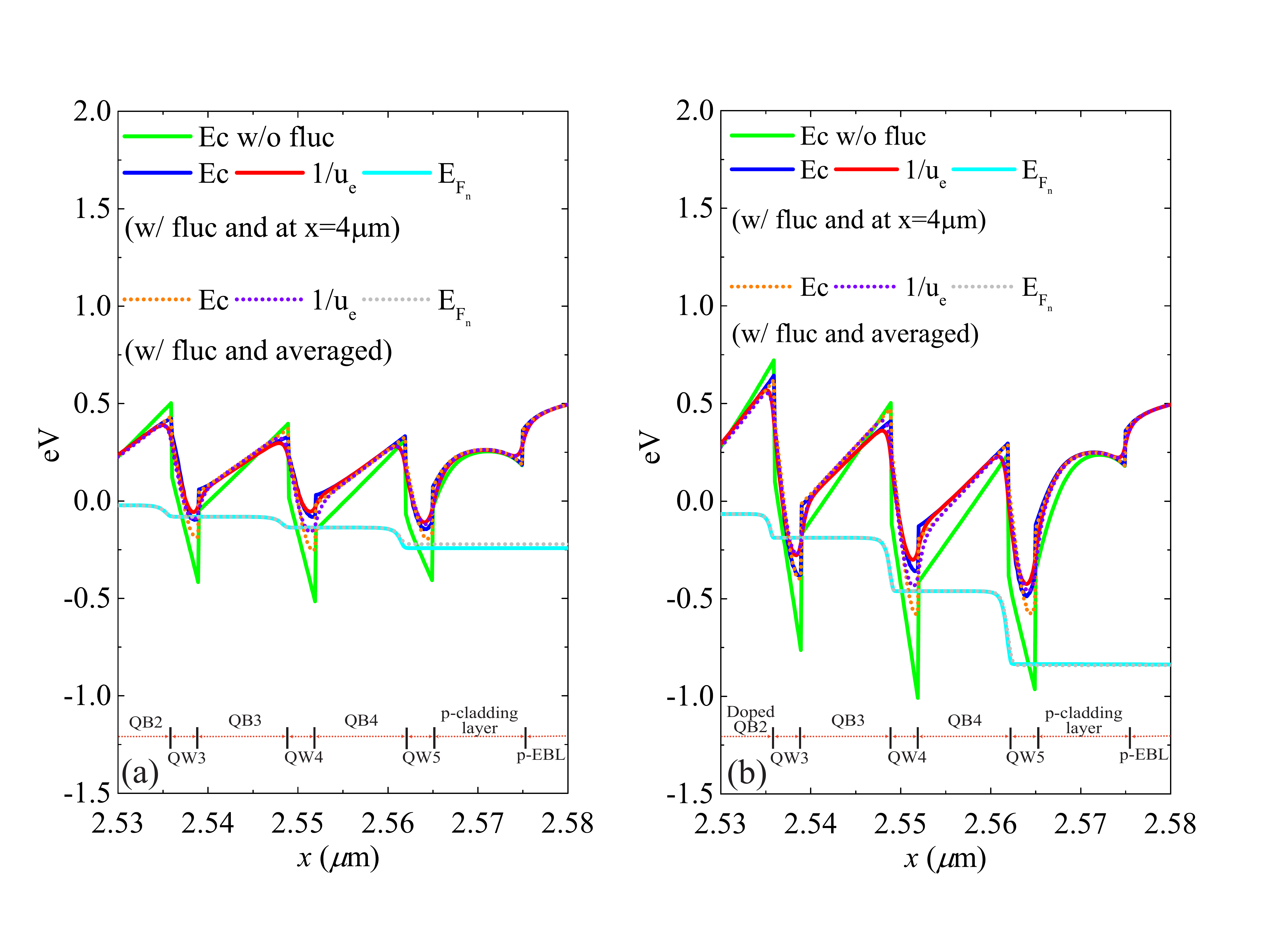}}}
\caption{Cross section plot for the three top QWs of $E_\text{c}$ without
fluctuations ($E_\text{c}$ w/o  fluc), $E_\text{c}$ with fluctuations ($E_\text{c}$ w/  fluc),effective potential $\frac{1}{u_\text{e}}$ , $E_{\text{F}n}$ along z-direction with random alloy fluctuations but without V-defects. (a) is for the blue LED, and (b) is for the green LED. The applied bias is 3V. The full curves w/ fluc are for the cross section with minimum barrier and the dashed lines are for the fluctuation-averaged potential $E_\text{c}$, $\frac{1}{u_\text{e}}$ and $E_{\text{F}n}$.}
\label{potential_compared}
\end{figure}

Figure \ref{potential_compared} shows the calculated $E_\text{c}$ and effective quantum potential $\frac{1}{u_\text{e}}$ for the blue and green LEDs, taking into account random alloy fluctuations at a bias voltage $V = 3$ V. A cross section with minimum potential barrier was chosen for the full curves. They show the best case for carrier transport across the QWs. The dashed curves for disorder averaged levels show slightly larger barriers. Without considering the random alloy fluctuation, the voltage voltage would be even higher which has been discussed in ref. \cite{lheureux20203d,yang2014influence,li2017localization}.Hence we will not discuss the cases with and without random alloy fluctuation. Although Figs. \ref{potential_compared}(a) and \ref{potential_compared}(b) show that random alloy fluctuations already change the depth of the QWs for blue LEDs (compare the $E_\text{c}$ curves with  and without fluctuations), the potential barrier induced by piezoelectric field between QWs is still about 0.4 eV for the blue LED, whose carriers can overcome the barrier through thermal excitation. However, for the green LED, the potential barrier is about 0.7 eV, which strongly impedes carrier transport across the MQWs at the bias $V = 3$ V. While the current density at that bias is a few A/cm$^2$ for the blue LED [see  Fig. \ref{blue_IQE_voltage}(a)], it is negligible in the green LED [Fig. \ref{green_IQE_voltage}(a)]. Thus, even when considering random alloy fluctuations, carriers are hardly injected directly into the $c$-plane QW for the green LED, despite the  large 0.7 V excess voltage. We also observe that the inclusion of the quantum disorder correction through the use of the LL theory (thus yielding the effective potential $\frac{1}{u_\text{e}}$) only weakly diminishes the potential barriers, in particular for the green LED. Reducing $V_\text{for}$ in these LEDs requires the use of another injection mechanism, namely, injection through V-defects.

\begin{figure}[tb]
\centerline{\scalebox{0.15}{\includegraphics{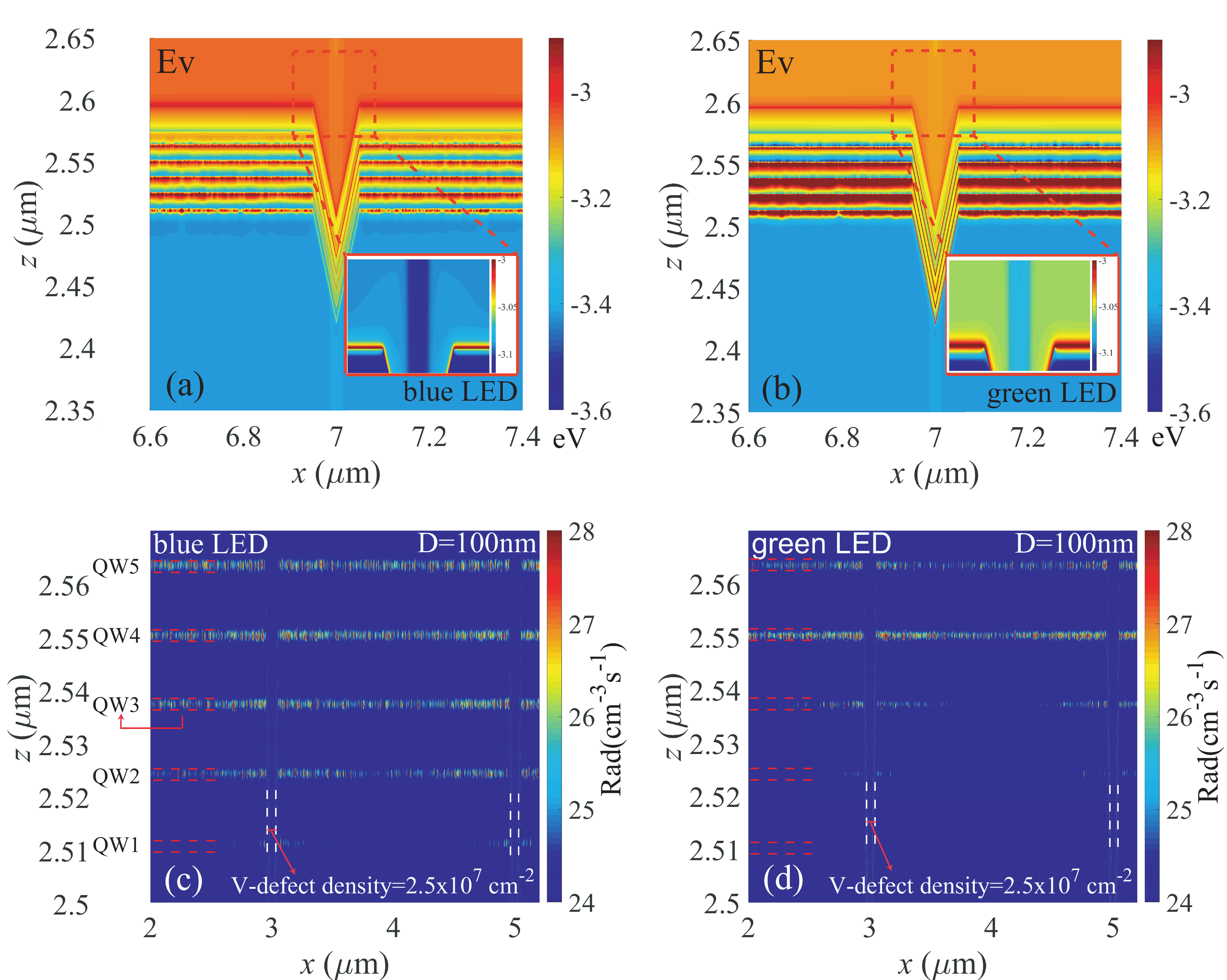}}}
\caption{Valence-band energy and radiative recombination for  blue LED and green LED at V-defect density $2.5\times$10$^7$ cm$^{-2}$ and $D = 100$ nm at a current density of 20 A/cm$^2$. (a) Valence band for the blue LED, (b) valence band for the green LED, (c) radiative recombination for the blue LED at 20 A/cm$^2$, and (d) radiative recombination for the green LED at 20 A/cm$^2$. The plot is at log scale. White dashed lines show the location of V-defects on the $x$ axis. Insets in panels (a) and (b) at different color scale emphasize the higher potential energy for holes at the center due to trap charging. The scale color bar ranges from -3.0eV (red) to -3.1 eV(blue)}
\label{VB_Rad}
\end{figure}

\begin{figure}[tb]
\centerline{\scalebox{0.15}{\includegraphics{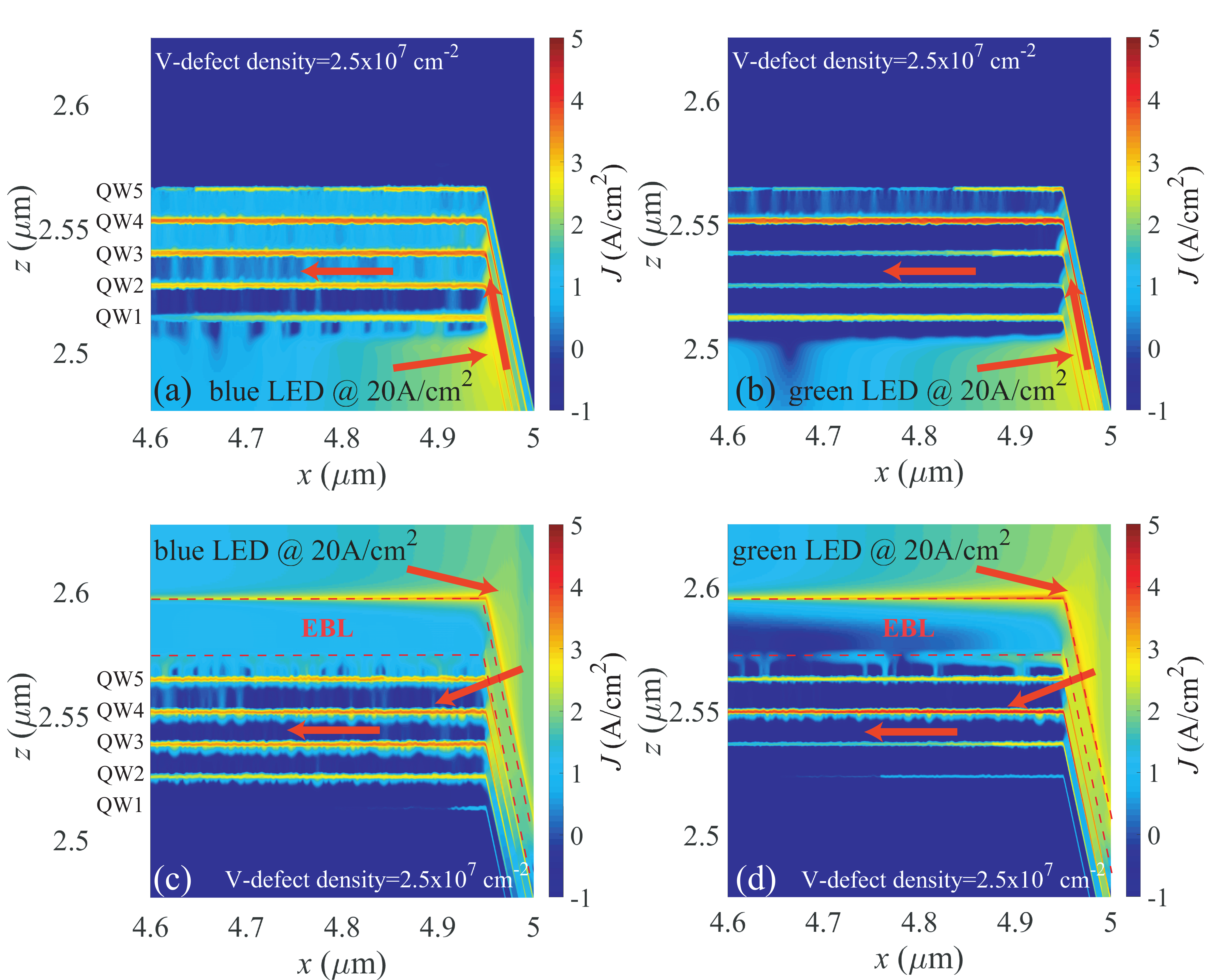}}}
\caption{Current distribution for  blue LED and green LED at V-defect density of $2.5\times$10$^7$ cm$^{-2}$ and $D = 100$ nm at a current density of 20 A/cm$^2$. (a) Electron current density |$J_n$| for blue LED, (b) electron current density |$J_n$| for green LED, (c) hole current density |$J_p$| for blue LED, and (d) hole current density |$J_p$| for green LEDs. The plot is at log scale. The vector sense for the electrons [panels (a) and (b)] and hole [panels (c) and (d)] are shown by the red arrows.}
\label{Jn_Jp}
\end{figure}

We first focus on hole injection through a V-defect. Figures \ref{VB_Rad}(a) and \ref{VB_Rad}(b) show the valence-band potential of blue and green LEDs in a V-defect, respectively. First, in the middle of the V-defect, within $\pm$15 nm of the TD in the $p$-GaN layer region, holes are trapped in the hole traps associated with the TD,  which repels holes from the TD center [the repelling potential is shown in a lighter color in the TD region in Figs. \ref{VB_Rad}(a) and \ref{VB_Rad}(b)]. Next, holes transport toward the $c$-plane QWs through the inclined plane, as shown in Figs. \ref{Jn_Jp}(c) and \ref{Jn_Jp}(d). Given the position of the electron-injecting region relative to the $c$-plane QWs, electrons are first injected into the side-wall QWs, from which they diffuse toward the lower-energy $c$-plane QWs.

Figures \ref{VB_Rad}(c) and \ref{VB_Rad}(d) show the radiative recombination map for the blue and green LEDs at 20 A/cm$^2$, respectively. For blue LEDs, recombination occurs throughout the structure as electrons and holes (at least four of the $c$-plane QWs for holes) are injected vertically [Figs. \ref{Jn_Jp}(a) and \ref{Jn_Jp}(c)] in the $c$-plane QWs. For the green LED, while electrons are injected both vertically and through the V-defect [Fig. \ref{Jn_Jp}(b)],  holes are not injected vertically because of the much larger polarization-induced barrier in the $c$ plane, Holes are therefore injected through the V-defect sidewall near which recombination occurs and decays laterally within a characteristic diffusion length. In addition, only two to three $c$-plane QWs receive holes through the V-defect. Both effects limit the effectiveness of V-defects as a solution to excite the whole volume of the $c$-plane QWs. For the blue LED, the crowding effect of carriers near  V-defects is weaker because holes can still   be injected vertically into the whole $c$-plane area due to random alloy fluctuations.  The relative importance of vertical injection versus V-defect injection is discussed in more detail below along with how the  V-defect size and density affects $V_\text{for}$ and the $IQE$ for blue and green LEDs.

A remarkable phenomenon is seen in Fig. \ref{Jn_Jp}: both electrons and holes are preferentially injected into QWs near the middle of the MQW stack,  QWS \#2, 3,4 for the blue LED, and QW \#4 for the green LED. This is dominantly due to the peculiarity of hole injection through V-defects. Due to the diminished piezoelectric induced barrier at the sidewall (which is a semipolar plane at 60 degree), the electrons and holes are more easily injected sideways than vertically, especially for holes, which have an heavier effective mass and lower mobility. Due to the p-doping of the last barrier, holes are easily injected in QW5 through the sidewall. Overcoming one more barrier is still relatively easy for holes, compared to the c-plane vertical injection. Holes then first meet electrons in QW4 and recombine there.  Why aren’t electrons injected in QW5? Close inspection of electron current in Fig.  \ref{Jn_Jp} (b) for green LEDs show that they experience a significant energy barrier between the sidewall QWs 4 and 5 (blue color of the barrier). This is due to the nearby p doping inside the V-defect which raises the barrier. The behavior is more significant for the green case as can be seen in Figs. \ref{Jn_Jp}(b) and (d), as well as on the recombination current in Fig. \ref{VB_Rad}(d). This was not observed without V-defect since it is much harder for holes to overcome the large piezoelectric barrier in the c-plane direction especially in the green case. Then, holes are preferentially injected in QW \# 5 where they meet electrons. It is well known that holes in the planar LED case are mainly injected in the QW nearest to the p-side in planar MQW LEDs, and this is of course well reproduced in our other simulation papers which include disorder effects without V-defects.

Since carriers are mainly going through V-defects for the green LED, the density of V-defects  play an important role in the LED current-voltage characteristics. Using a high V-defect density to obtain more injection sites, the active $c$-plane QW area  becomes smaller and the carrier density in the $c$ plane   increases for the same current. This  adversely affects the $IQE$ and droop behavior. \cite{ryu2009rate} Therefore, it is important to discuss in detail how the V-defect density and size affect the LED structure.

\subsection{Influence of V-defects and random alloy fluctuations on  $I$-$V$ and $IQE$ }

The influence of  V-defects for the specific lateral LED of Fig. \ref{lateral LED and V-pit definition} is now discussed for green and blue LEDs. The effects on $IQE$ and $V_\text{for}$ of both random alloy fluctuations and V-defects are simulated for different V-defect densities and diameters. We use V-defect densities  of  $1\times 10 ^6$, $2.5\times 10 ^7$, $1\times 10 ^8$, $2.25\times 10 ^8$, and $6.25\times 10 ^8$~cm$^{-2}$ and V-defect diameters of 50, 100, 220, 280, and 340 nm.
\subsubsection{Performance of blue V-defect  LED}

\begin{figure}[tb]
\centerline{\scalebox{0.15}{\includegraphics{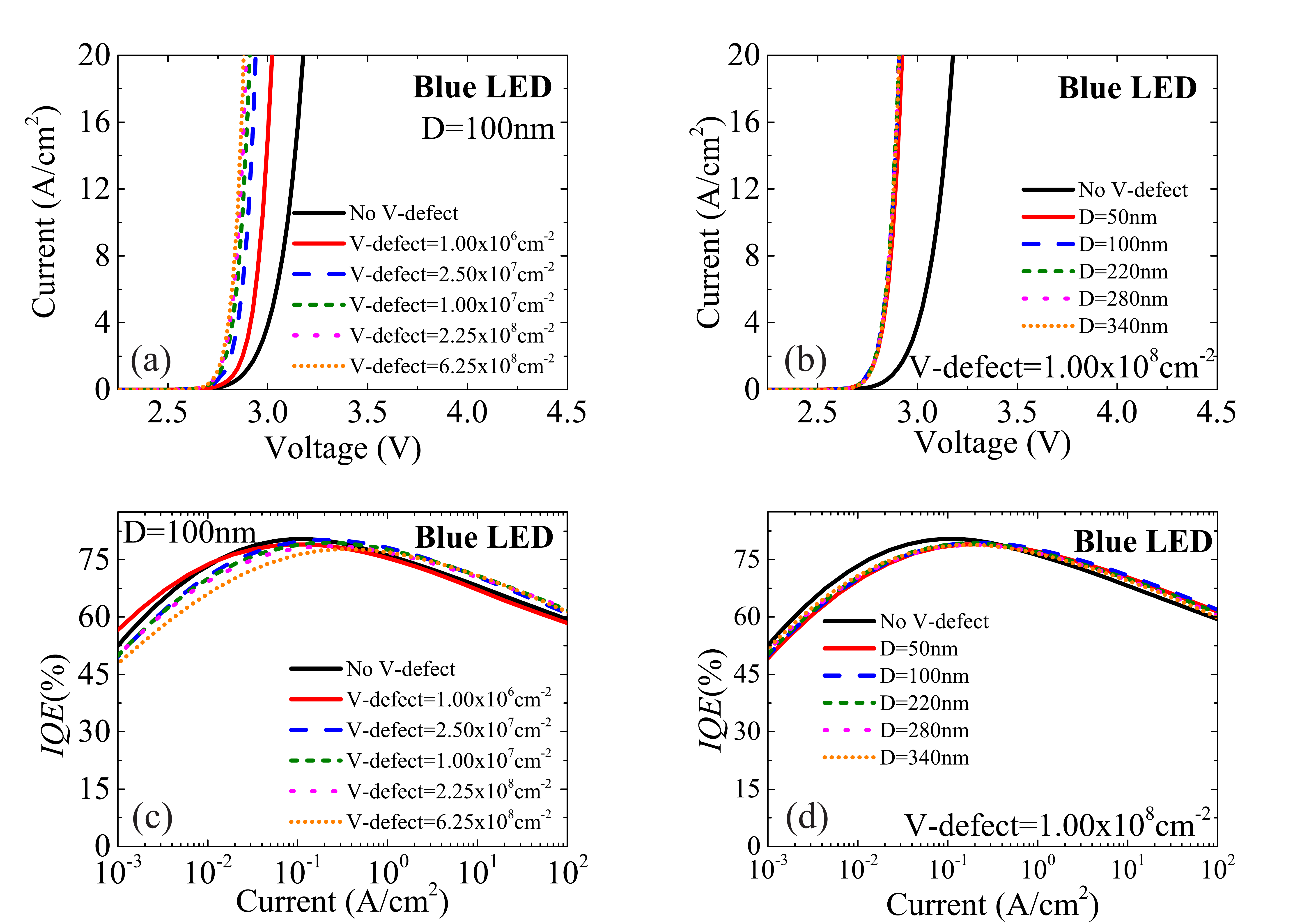}}}
\caption{(a) Voltage versus current for blue LED at different V-defect densities for V-defect diameter $D = 100$ nm. (b) Voltage versus current for blue LED at different V-defect diameters and a V-defect density of $1\times 10 ^8$~cm$^{-2}$. (c)  $IQE$ versus current for blue LED at different V-defect densities and a V- defect diameter of $D = 100$ nm. (d)  $IQE$ versus current for blue LED at different V-defect diameters and a V-defect density of $1\times 10 ^8$~cm$^{-2}$.}
\label{blue_IQE_voltage}
\end{figure}

We first discuss the dependence of $V_\text{for}$ on the V-defect density. Here we define the forward voltage $V_\text{for}$ as the voltage needed for a current density of 1 A/cm$^2$. Figure \ref{blue_IQE_voltage}(a) shows the results for the blue LED structures without and with the V-defects. The V-defect diameter is 100 nm. $V_\text{for}$ is about 2.88 V for the case without V-defects. The maximum density of V-defects is $6.25\times 10 ^8$~cm$^{-2}$, and  $V_\text{for}$ for this case is about 2.74~V at 1 A/cm$^2$, which equates to about zero excess voltage. 

For a V-defect density of  $1\times 10 ^8$~cm$^{-2}$, Fig. \ref{blue_IQE_voltage}(b) shows the dependence on V-defect size. The results show that the voltage  decreases as the V-defect diameter increases from 50  to 100 nm, but $V_\text{for}$  remains constant for the larger V-defect diameters, with all the decrease in $V_\text{for}$ already realized at the lateral V-defect size of 50 nm. One factor contributing to this result is  that the voltage is already close to $\frac{h\nu}{e}$, leaving little possibility for  further reduction of $V_\text{for}$. 

Figure \ref{blue_IQE_voltage}(c) shows the $IQE$ of blue LEDs  with different V-defect densities; the $IQE$ peak  slightly decreases as the V-defect density increases. For high V-defect densities, the $IQE$ peak appears at higher current density. In the low-current-density region (10$^{-3}$ to 1.0 A/cm$^2$), holes are injected mainly through  V-defects and electrons are injected vertically through the random alloy fluctuating potential, whereas for higher current density,  the higher $V_\text{for}$ means that both carriers start to be injected directly from the $c$-plane QW through the fluctuating potential barriers. Since the V-defect center contains strong nonradiative recombination (NR) centers,  carriers are injected through V-defects experience more NR. 

\begin{figure}[tb]
\centerline{\scalebox{0.15}{\includegraphics{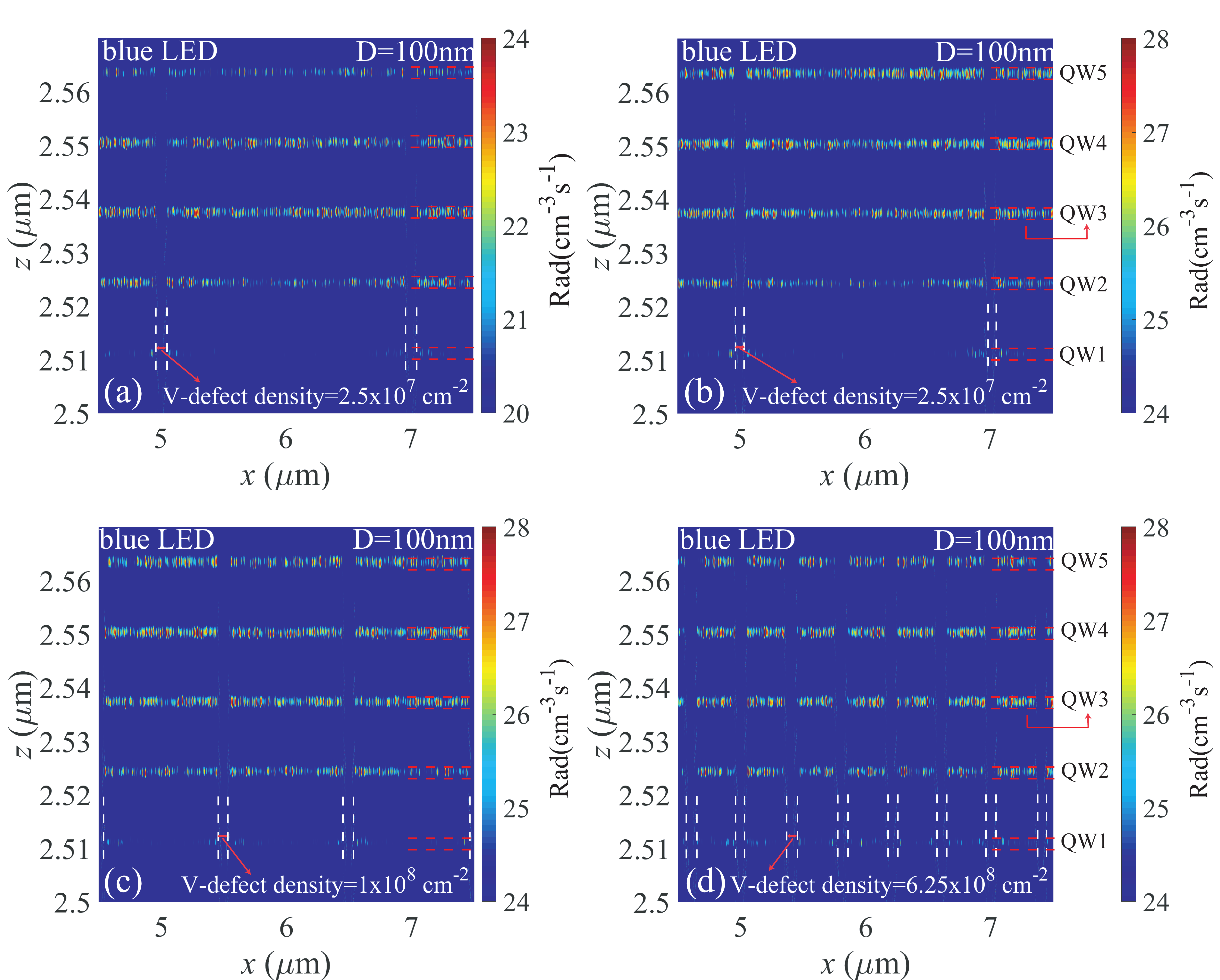}}}
\caption{Distribution of  radiative recombination rate for  blue LED with $D = 100$ nm. The plot is at log scale. (a) V-defect density $2.5\times 10 ^7$~cm$^{-2}$ at current density 0.01 A/cm$^2$ (b) V-defect density $2.5\times 10 ^7$~cm$^{-2}$ at current density 20 A/cm$^2$ (c) V-defect density $1\times 10 ^8$~cm$^{-2}$ at current density 20 A/cm$^2$ (d) V-defect density $6.25\times 10 ^8$~cm$^{-2}$ at current density 20 A/cm$^2$. White dashed lines indicate the location of V-defects on $x$ axis. The red arrow in panel (b) indicates that the top QW is not injected vertically at low current (bias) and injection occurs through the V-defects. The red dashed lines indicate the QW positions.}
\label{blue_rad}
\end{figure}

Figures \ref{blue_rad}(a) and \ref{blue_rad}(b) show the radiative recombination for current densities of 10$^{-2}$  and 20 A/cm$^2$, respectively, for blue LEDs at a V-defect density of $2.5\times 10 ^7$~cm$^{-2}$. At low current density, the top QW shows weak radiative recombination, which means that holes are injected from V-defects due to the diminished piezoelectric field-induced barrier at the sidewall, especially for holes with lower mobility. At high current density, the top QW also emits light because holes are also injected vertically. Thus, at higher current density, the $IQE$ and $V_\text{for}$ are less influenced by V-defects and the dislocation region. More QWs emit light, and the impact of NR diminishes, so all curves regroup. 

Figures \ref{blue_rad}(b)--\ref{blue_rad}(d) show the recombination distribution for different V-defect densities at $J = 20$ A/cm$^2$. The crowding effect is limited because of vertical injection through random alloy fluctuations but is significant at low V-defect density. Note that if  random alloy fluctuations are not considered in the c-plane QW, carrier diffusion extends much further. Then, we do not observe current crowding near V-defect for the lower V-defect density, making the predicted $IQE$ higher. Increasing the V-defect density diminished the crowding effect and thus moves the peak $IQE$ to a higher current density, although the peak $IQE$  is diminished somewhat because of the increased density of NR centers [Fig. \ref{blue_IQE_voltage}(c)]. Figure \ref{blue_IQE_voltage}(d) shows the $IQE$ for different V-defect diameters at a V-defect density of $1.0\times 10 ^8$~cm$^{-2}$. When the diameter increases, the V-defect area increases and the active $c$-plane area decreases. Thus, the $IQE$ decreases with increasing V-defect diameter  because of the increased Auger effect, which is consistent with earlier work. \cite{li20163d}

\begin{figure}[tb]
\centerline{\scalebox{0.15}{\includegraphics{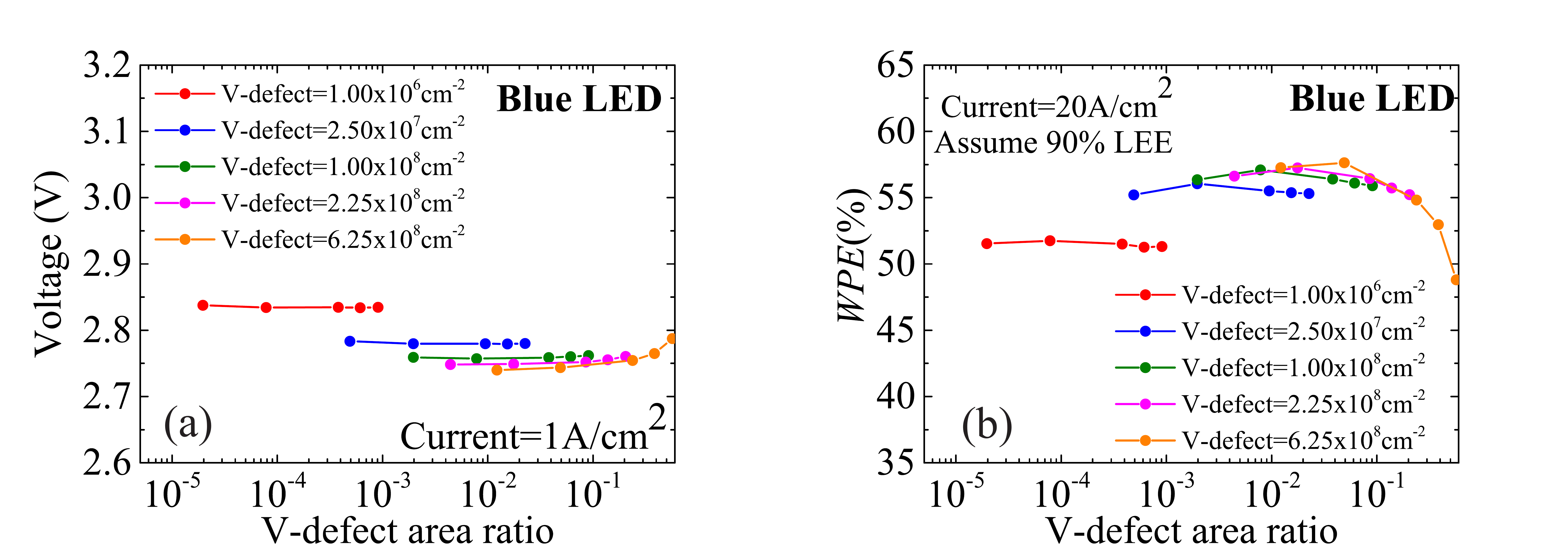}}}
\caption{(a)  V-defect region ratio (V-defect area/chip area) versus turn-on voltage for different V-defect densities at a current density of 1 A/cm$^2$ for blue LEDs. (b)  V-defect region ratio versus $WPE$ for different V-defect densities at a current density of 20 A/cm$^2$ for blue LEDs. The 90\% LEE is assumed.}
\label{blue_ratio}
\end{figure}

To further to discuss the joint influence of V-defect density and diameter on the $IQE$, we  consider the ratio of V-defect area to the total chip area. Figure \ref{blue_ratio}(a) shows  $V_\text{for}$ as a function of the V-defect area ratio, where  points on a given V-defect density curve correspond to the different V-defect diameters (50, 100, 220, 280, and 340 nm), which change the V-defect area ratio. At low V-defect area ratio, $V_\text{for}$ is affected by both the V-defect area ratio \cite{wang2019investigating} and the V-defect density. This result is attributed to the density being too low, even for large V-defects, so the  carriers injected from V-defects need to travel through a long lateral region to inject into the entire QW. In addition, the natural alloy potential fluctuations in the QW limit the lateral transport distance of electrons and holes. As a result, carriers become crowded near the V-defect at low V-defect density, so the device performance is strongly limited by the lateral transport distance from the V-defect within the QW. \cite{shen2021three} 

This result is again   verified by Figs. \ref{blue_rad}(b)--\ref{blue_rad}(d). The recombination becomes less crowded when the V-defect density increases. Once the V-defect area ratio grows larger, $V_\text{for}$ decreases and then saturates at the expected value near 2.7 V. Another contribution comes from the modified vertical injection due to the presence of V-defects: the  fluctuations in the vertical energy barriers induced by the random alloy fluctuations causes some carriers to be directly injected from the $c$ plane in regions with lower barriers. With V-defects, electrons and holes can flow laterally into all quantum wells. \cite{li2019nanoscale} Once carriers flow into QWs, they screen the polarization-related electric fields so that vertical current injection directly into $c$-plane QWs increases at lower bias voltage compared with the case without V-defects. 

The diameter and density of V-defects also affect the efficiency of the blue LED. Since the optimized performance is decided by wall plug efficiency($WPE$), we estimate the $WPE$ by assuming a 90\% light extraction efficiency (LEE), typical of LEDs grown on patterned sapphire substrates(PSS). We simulated how the $WPE$ at a fixed 20 A/cm$^2$ current density depends on the V-defect area ratio, which corresponds to different diameters and densities of V-defects. The results are shown in Fig. \ref{blue_ratio}(b). Interestingly,  the low V-defect density  of $1.0\times 10 ^6$ and $2.5\times 10 ^7$~cm$^{-2}$ have a much lower $WPE$ and $IQE$ than the others because the current paths from V-defects to the whole chip area are too long so the constant injected current of 20 A/cm$^2$ becomes crowded near the V-defect region. They also have a higher $V_\text{for}$. When the V-defect density exceeds $1.00\times 10 ^8$~cm$^{-2}$, which is within the lateral diffusion length, the carriers are easily injected into the whole active region, diminishing crowding and further canceling the quantum-confined Stark effect. This is also seen in Figs. \ref{blue_rad}(b)--\ref{blue_rad}(d). 

This latter effect helps to increase electron-hole overlap and increase the efficiency. However, the $IQE$  also gradually decreases as the V-defect area ratio further increases because the active volume decreases, and the droop effect starts to dominate at lower current density for larger V-defect area ratio. The conclusion is obvious: it is advantageous (i) to have a high V-defect density so that the V-defect separation is about the lateral carrier transport distance (often referred to as a diffusion length), to inject carriers homogeneously, and (ii) to have small V- defects to minimize nonemitting, unproductive V-defects. The larger V-defects densities also have a smaller $V_\text{for}$, but the reduction of $V_\text{for}$ saturates as V-defects density becomes larger than $10^{8}$ cm$^{-2}$. Hence, the $WPE$ reaches a peak at the V-defect ratio around 5\%. 

\subsubsection{ Performance of green  V-defect  LED}

\begin{figure}[tb]
\centerline{\scalebox{0.15}{\includegraphics{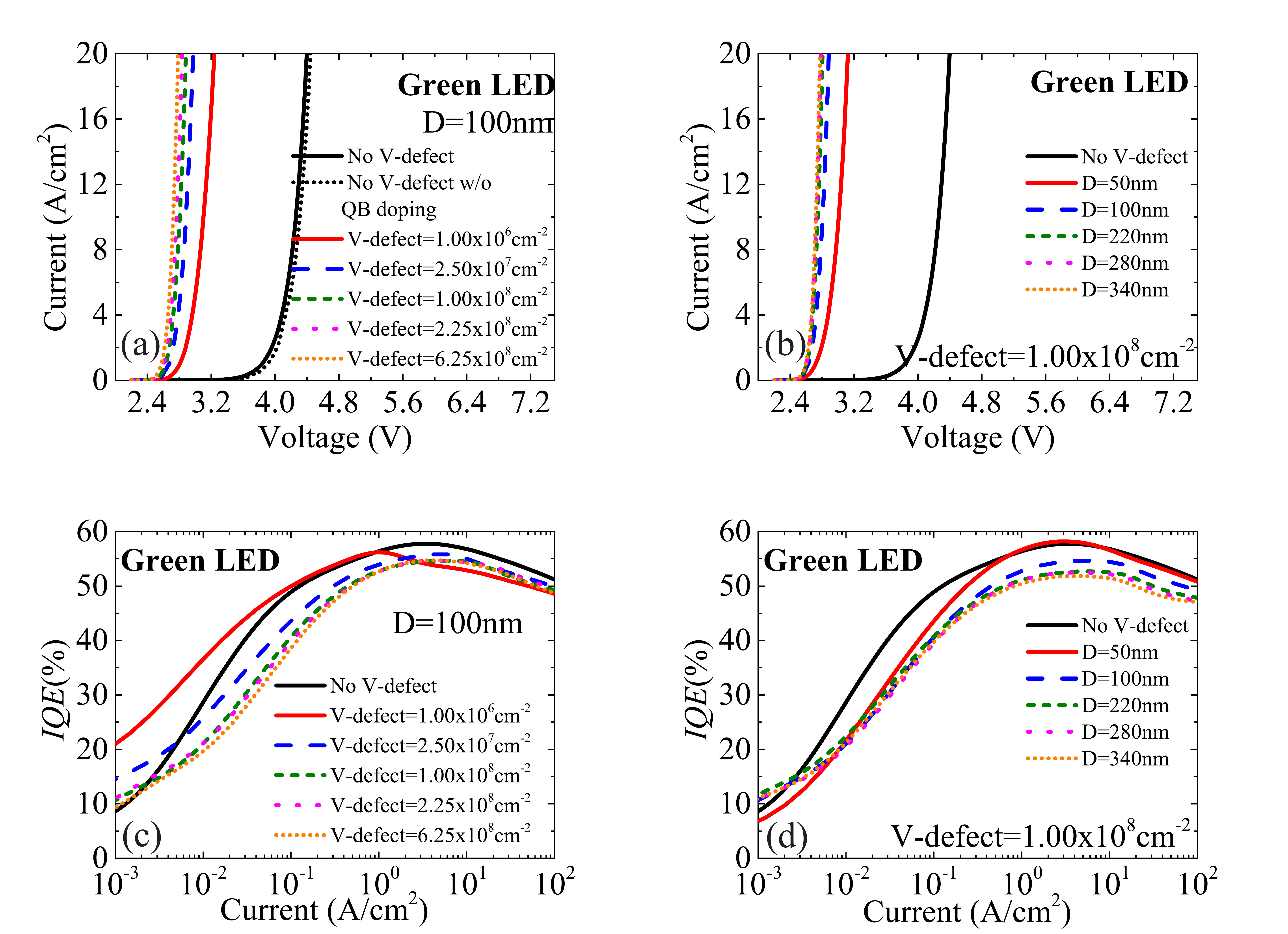}}}
\caption{(a) Voltage versus current for green LED at different V-defect densities and for a V-defect diameter of $D = 100$ nm. (b) Voltage versus current for green LED at different V-defect diameters and a V-defect density of $1\times 10 ^8$~cm$^{-2}$. (c)  $IQE$ versus current density for green LED at different V-defect densities and for a V-defect diameter of $D = 100$ nm. (d)  $IQE$ versus current for green LED at different V-defect diameters and a V-defect density of $1\times 10 ^8$~cm$^{-2}$.}
\label{green_IQE_voltage}
\end{figure}

For the green LED without V-defects,  $V_\text{for}=3.83$ V at 1 A/cm$^2$ current density is greater than for the blue LED  [Fig. \ref{green_IQE_voltage}(a)]. \cite{lheureux20203d,lynsky2020barriers} If we further remove the barrier doping, as indicated by  the black dotted line,  $V_\text{for}$ increases. Despite doping the QB barriers between the MQWs, the polarization barrier induced by the high indium composition is not completely offset. Therefore, V-defects are much more helpful to decrease the turn-on voltage for green LEDs than for blue LEDs. The lowest V-defect density  of $1\times 10 ^6$~cm$^{-2}$  decreases the turn-on voltage to 2.81 V, as shown in Fig. \ref{green_IQE_voltage}(a). 

We also simulated different V-defect densities and find that $V_\text{for}$ decreases as the V-defect density increases. The different diameters of V-defects are also simulated at a density $1\times 10 ^8$~cm$^{-2}$, as shown in Fig. \ref{green_IQE_voltage}(b). The results are similar to the blue LED shown in Fig. \ref{blue_IQE_voltage}(b). The voltage decreases as the diameter increases and then saturates.

We also calculate  $IQE$ versus V-defect density  for  $D = 100$ nm [see Fig. \ref{green_IQE_voltage}(c)]. The results show that different V-defect densities produce different performances. Unlike the blue LED where some carriers are injected from the $c$ plane, in the green LED, most carriers are injected into the planar QWs through the V-defect sidewall, as shown in Figs. \ref{VB_Rad} and \ref{Jn_Jp}. For the lower V-defect density, the $IQE$ peaks  slightly higher due to a lower TD density. However, due to greater current crowding, where carriers are crowded in the $c$-plane QW near V-defects, the droop effect  appears at lower current densities, as shown in Fig. \ref{green_IQE_voltage}(c). When the current density increases, additional carriers  flow into the whole active QW region for all V-defect densities. Therefore, the $IQE$s reach a similar value for all V-defect densities, as shown in Fig. \ref{green_IQE_voltage}(c). 

Figure \ref{green_IQE_voltage}(d) shows the $IQE$ for different V-defect diameters and for a V-defect density of $1.0\times 10 ^8$~cm$^{-2}$. For the larger diameters, the V-defect area  increases and the active $c$-plane area  decreases. Thus, the $IQE$ decreases as the V-defect diameter increases. To further discuss this, we consider how the performance depends on the V-defect area ratio. 

\begin{figure}[tb]
\centerline{\scalebox{0.15}{\includegraphics{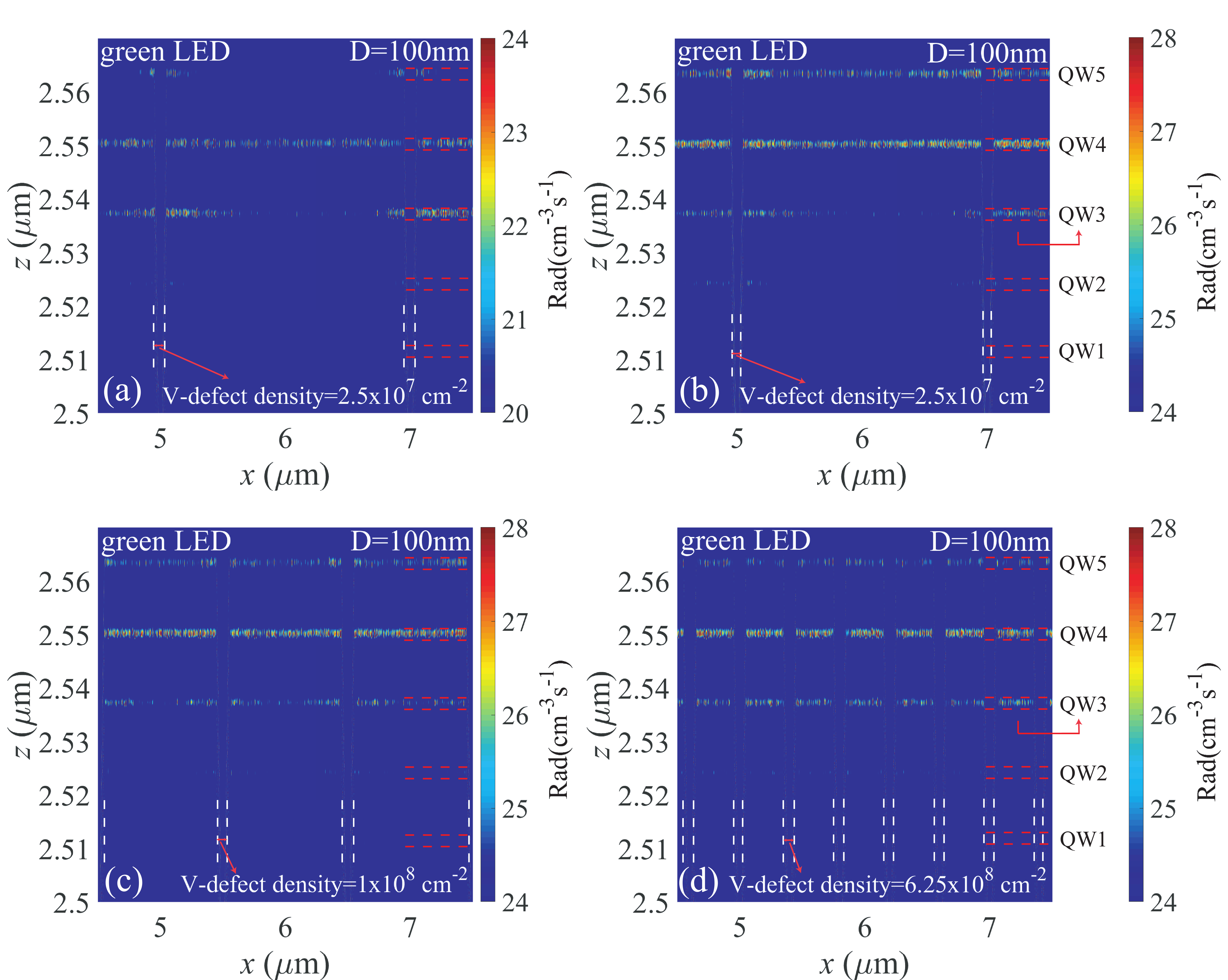}}}
\caption{Distribution of  radiative recombination rate for  green LED at $D = 100$ nm. The plot is at log scale.(a) V-defect density $2.5\times 10 ^7$~cm$^{-2}$ at current density 0.01 A/cm$^2$. (b) V-defect density $2.5\times 10 ^7$~cm$^{-2}$ at current density 20 A/cm$^2$. (c) V-defect density $1\times 10 ^8$~cm$^{-2}$ at current density 20 A/cm$^2$. (d) V-defect density $6.25\times 10 ^8$~cm$^{-2}$ at current density 20 A/cm$^2$. Vertical white dashed lines indicate the location of V-defects along the $x$ axis.}
\label{green_rad}
\end{figure}

Figure \ref{green_rad} shows the recombination distribution for different current densities and different V-defect densities. Unlike for blue LEDs, a strong crowding effect occurs until the V-defect density exceeds $1\times 10 ^8$~cm$^{-2}$, independently of the  current density. This again shows that the V-defect density is the dominant factor for carrier injection. As mentioned in the blue case, if the random alloy fluctuation is not considered, the current crowding effect is not observed, which leads to a much higher $IQE$ and the delay of droop behavior.

\begin{figure}[tb]
\centerline{\scalebox{0.15}{\includegraphics{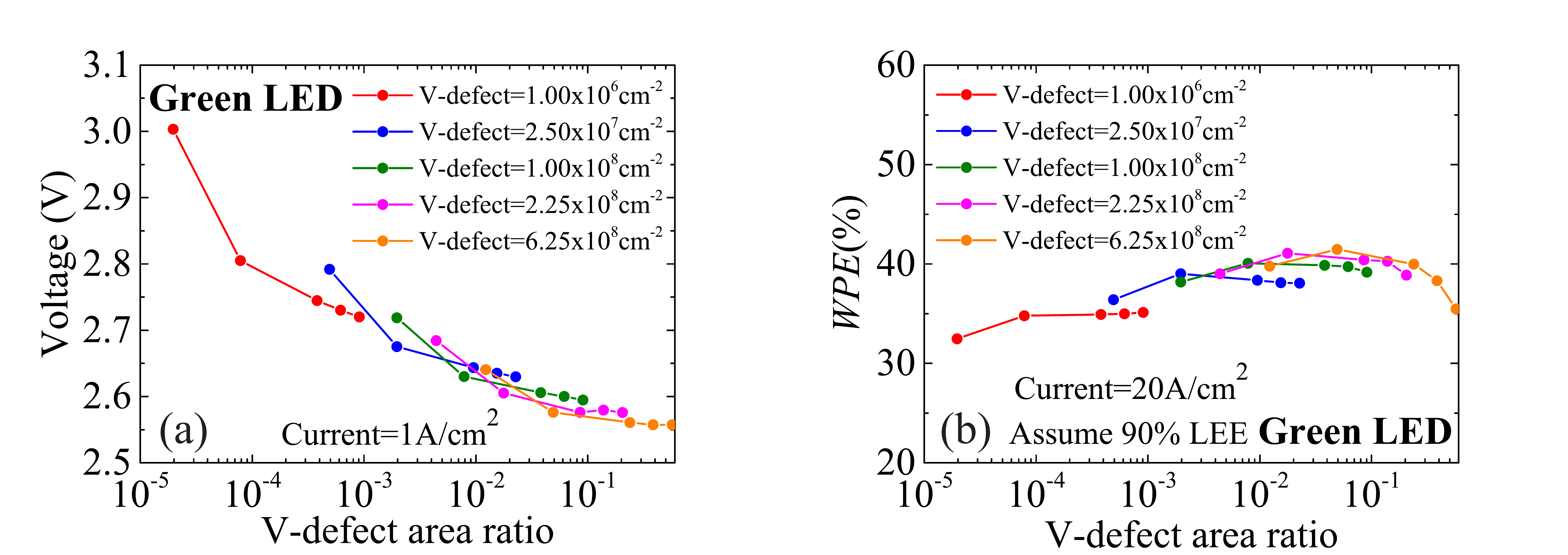}}}
\caption{(a)  V-defect region ratio (V-defect area/chip area)  versus turn-on voltage for different V-defect densities at a current density 1 A/cm$^2$ for green LED. (b)  V-defect region ratio versus $WPE$ for different V-defect densities at a current density of 20 A/cm$^2$ for green LED. The 90\% LEE is assumed.}
\label{green_ratio}
\end{figure}

Figure \ref{green_ratio}(a) shows $V_\text{for}$ versus the V-defect area ratio and for different V-defect densities.  Figure \ref{green_ratio}(a) shows that the turn-on voltage for green LEDs decreases as the V-defect area ratio increases. In addition, green LEDs have a higher indium composition and a higher polarization-related electric field, so only holes   flow into the $c$ plane from V-defects and become crowded at the last QWs, as do  electrons [see Fig. \ref{Jn_Jp}(b)]. Therefore, the turn-on voltage  decreases as the V-defect area ratio increases. This effect influences $WPE$ of the green LED. 

Figure \ref{green_ratio}(b) shows how the V-defect area ratio affects the $WPE$ at 20 A/cm$^2$. Again a 90\% LEE is assumed in estimating the $WPE$. As mentioned, $V_\text{for}$ decreases as the V-defect area ratio increases, which improve the $WPE$. At the same V-defect area ratio, a greater V-defect density corresponds to a greater $IQE$. Because  carriers are injected into the QW from V-defects and the diffusion length in the fluctuating QWs is short, the $IQE$ depends much more on the V-defect density than on the V-defect diameter. Combining both effects, the $WPE$ reaches a maximum near 5\% V-defect area ratio. For the same V-defect area ratio, a high defect density is preferable in green LEDs since they provide a higher $IQE$ for a similar $V_\text{for}$.

%Figs. \ref{blue_ratio}(b) abd\ref{green_ratio}(b) shows estimated wall plug efficiency by assuming light extraction efficiency to be 90 \% at 20 A/cm$^{2}$. As shown in the data, the voltage is higher with a smaller V-defect density where the $WPE$ is lower. As the V-defect density increases, the $WPE$ improves due to the decrease of $V_\text{for}$. Then it reaches a peak at the V-defect ratio around $~$5\%. As the V-defect ratio increases over 10 \%, the influence of nonradiative recombination at the V-defect center and relatively smaller active volumes start to dominate because the reduction of voltage is saturated.

%
%\begin{figure}[htpb]
%\centerline{\scalebox{0.30}{\includegraphics{fig12.pdf}}}
%\caption{(a)  The $WPE$ versus V-defect region ratio for different V-defect densities at a current density 20 A/cm$^2$ for blue LED. (b) The $WPE$ versus V-defect region ratio for different V-defect densities at a current density 20 A/cm$^2$ for green LED. The light extraction efficiency is not not considered since this work does not discuss light extraction issue.}
%\label{WPE_BLUE_GREEN}
%\end{figure}

\section{Conclusion}

In contrast with previous analyses of V-defects \cite{li20163d} in LEDs, we computed  the influence of both random alloy fluctuations and of V-defect density and size. Random alloy fluctuations help to inject carriers  vertically, whereas weak polarization-related electric fields in the inclined sidewalls of V-defects provide an alternative path for lateral carrier injection into the $c$-plane QWs. In blue LEDs,  V-defects have only a small impact, mainly reducing $V_\text{for}$ by 0.14 V because carriers are rather efficiently injected vertically. This feature cannot be observed without including both random alloy fluctuaion and V-defect. With considering only the V-defect model\cite{li20163d}, carriers can only diffuse into QW through the V-defect where the influence of V-defect density might be overestimated. In green LEDs,  the large polarization-related barriers hamper vertical transport, so carriers are preferentially injected from V-defects. $V_\text{for}$ is strongly reduced by about 1.0 V, even at small V-defect density, but the limited lateral diffusion length induces current crowding near V-defects. Therefore, increasing V-defect density increases the peak $IQE$ at large current densities. With the use of an alternative 2D model to 3D models, we could simulate a much larger, realistic area of LEDs and observe the limits of lateral carrier diffusion from V-defects due to random alloy fluctuations, while these features cannot be seen in 3D simulations as the modeled LED size is well below the diffusion length, severely limited by computing resources. 

When carriers are injected into the $c$-plane QWs at high bias and current density, they can screen the polarization field and thereby allow more carriers to be injected either through V-defects or vertically through the $c$-plane stack of barriers and wells, thus providing a welcome synergy between V-defects and vertical carrier injection. However, optimization will require further work. Although the increase of V-defect density decreases $V_\text{for}$, it also reduces the $IQE$ due to a higher possibility to be trapped and recombine nonradiatively  in  TDs around the center of V-defects. Detailed measurements of these nonradiative parameters are required for precise selection of V-defect density and size.
\section*{Acknowledgments}

This work was supported by the Ministry of Science and Technology, Taiwan (Grant Nos. MOST 108-2628-E-002-010-MY3, MOST 110-2923-E-002-002, and MOST 111-2923-E-002-009). The work at UCSB was supported by the National Science Foundation (Grant No. DMS-1839077), and by grants from the Simons Foundation (Grant Nos. 601952 to J.S. and 601954 to C.W.). This work was also supported by the TECCLON project, grant ANR-20-CE05-0037-01 of the French Agence Nationale de la Re-cherche (ANR). Additional support for JSS was provided by the U.S. Department of Energy (Award No. DE-EE0009691). The data used in this paper can be requested by contacting the corresponding author. The modeling tool was developed by the NTU laboratory, and more detailed information can be found on our website (http://yrwu-wk.ee.ntu.edu.tw/).

%\nocite{*}
\normalem
\bibliography{V-defect}% Produces the bibliography via BibTeX.

%aipnum4-2.bst 2019-01-14 (MD) hand-edited version of apsrev4-1.bst
\begin{thebibliography}{47}%
\makeatletter
\providecommand \@ifxundefined [1]{%
 \@ifx{#1\undefined}
}%
\providecommand \@ifnum [1]{%
 \ifnum #1\expandafter \@firstoftwo
 \else \expandafter \@secondoftwo
 \fi
}%
\providecommand \@ifx [1]{%
 \ifx #1\expandafter \@firstoftwo
 \else \expandafter \@secondoftwo
 \fi
}%
\providecommand \natexlab [1]{#1}%
\providecommand \enquote  [1]{``#1''}%
\providecommand \bibnamefont  [1]{#1}%
\providecommand \bibfnamefont [1]{#1}%
\providecommand \citenamefont [1]{#1}%
\providecommand \href@noop [0]{\@secondoftwo}%
\providecommand \href [0]{\begingroup \@sanitize@url \@href}%
\providecommand \@href[1]{\@@startlink{#1}\@@href}%
\providecommand \@@href[1]{\endgroup#1\@@endlink}%
\providecommand \@sanitize@url [0]{\catcode `\\12\catcode `\$12\catcode
  `\&12\catcode `\#12\catcode `\^12\catcode `\_12\catcode `\%12\relax}%
\providecommand \@@startlink[1]{}%
\providecommand \@@endlink[0]{}%
\providecommand \url  [0]{\begingroup\@sanitize@url \@url }%
\providecommand \@url [1]{\endgroup\@href {#1}{\urlprefix }}%
\providecommand \urlprefix  [0]{URL }%
\providecommand \Eprint [0]{\href }%
\providecommand \doibase [0]{https://doi.org/}%
\providecommand \selectlanguage [0]{\@gobble}%
\providecommand \bibinfo  [0]{\@secondoftwo}%
\providecommand \bibfield  [0]{\@secondoftwo}%
\providecommand \translation [1]{[#1]}%
\providecommand \BibitemOpen [0]{}%
\providecommand \bibitemStop [0]{}%
\providecommand \bibitemNoStop [0]{.\EOS\space}%
\providecommand \EOS [0]{\spacefactor3000\relax}%
\providecommand \BibitemShut  [1]{\csname bibitem#1\endcsname}%
\let\auto@bib@innerbib\@empty
%</preamble>
\bibitem [{\citenamefont {Weisbuch}(2019)}]{weisbuch2019search}%
  \BibitemOpen
  \bibfield  {author} {\bibinfo {author} {\bibfnamefont {C.}~\bibnamefont
  {Weisbuch}},\ }\bibfield  {title} {\enquote {\bibinfo {title} {{On the search
  for efficient solid state light emitters: Past, present, future}},}\
  }\href@noop {} {\bibfield  {journal} {\bibinfo  {journal} {ECS Journal of
  Solid State Science and Technology}\ }\textbf {\bibinfo {volume} {9}},\
  \bibinfo {pages} {016022} (\bibinfo {year} {2019})}\BibitemShut {NoStop}%
\bibitem [{\citenamefont {Auf~der Maur}\ \emph {et~al.}(2016)\citenamefont
  {Auf~der Maur}, \citenamefont {Pecchia}, \citenamefont {Penazzi},
  \citenamefont {Rodrigues},\ and\ \citenamefont
  {Di~Carlo}}]{PhysRevLett.116.027401}%
  \BibitemOpen
  \bibfield  {author} {\bibinfo {author} {\bibfnamefont {M.}~\bibnamefont
  {Auf~der Maur}}, \bibinfo {author} {\bibfnamefont {A.}~\bibnamefont
  {Pecchia}}, \bibinfo {author} {\bibfnamefont {G.}~\bibnamefont {Penazzi}},
  \bibinfo {author} {\bibfnamefont {W.}~\bibnamefont {Rodrigues}},\ and\
  \bibinfo {author} {\bibfnamefont {A.}~\bibnamefont {Di~Carlo}},\ }\bibfield
  {title} {\enquote {\bibinfo {title} {Efficiency drop in green
  $\mathrm{InGaN}/\mathrm{GaN}$ light emitting diodes: The role of random alloy
  fluctuations},}\ }\href {https://doi.org/10.1103/PhysRevLett.116.027401}
  {\bibfield  {journal} {\bibinfo  {journal} {Phys. Rev. Lett.}\ }\textbf
  {\bibinfo {volume} {116}},\ \bibinfo {pages} {027401} (\bibinfo {year}
  {2016})}\BibitemShut {NoStop}%
\bibitem [{\citenamefont {Auf~der Maur}\ \emph {et~al.}(2014)\citenamefont
  {Auf~der Maur}, \citenamefont {Barettin}, \citenamefont {Pecchia},
  \citenamefont {Sacconi},\ and\ \citenamefont {Di~Carlo}}]{6935331}%
  \BibitemOpen
  \bibfield  {author} {\bibinfo {author} {\bibfnamefont {M.}~\bibnamefont
  {Auf~der Maur}}, \bibinfo {author} {\bibfnamefont {D.}~\bibnamefont
  {Barettin}}, \bibinfo {author} {\bibfnamefont {A.}~\bibnamefont {Pecchia}},
  \bibinfo {author} {\bibfnamefont {F.}~\bibnamefont {Sacconi}},\ and\ \bibinfo
  {author} {\bibfnamefont {A.}~\bibnamefont {Di~Carlo}},\ }\bibfield  {title}
  {\enquote {\bibinfo {title} {{Effect of alloy fluctuations in InGaN/GaN
  quantum wells on optical emission strength}},}\ }in\ \href
  {https://doi.org/10.1109/NUSOD.2014.6935331} {\emph {\bibinfo {booktitle}
  {Numerical Simulation of Optoelectronic Devices, 2014}}}\ (\bibinfo {year}
  {2014})\ pp.\ \bibinfo {pages} {11--12}\BibitemShut {NoStop}%
\bibitem [{\citenamefont {Lynsky}\ \emph {et~al.}(2020)\citenamefont {Lynsky},
  \citenamefont {Alhassan}, \citenamefont {Lheureux}, \citenamefont {Bonef},
  \citenamefont {DenBaars}, \citenamefont {Nakamura}, \citenamefont {Wu},
  \citenamefont {Weisbuch},\ and\ \citenamefont {Speck}}]{lynsky2020barriers}%
  \BibitemOpen
  \bibfield  {author} {\bibinfo {author} {\bibfnamefont {C.}~\bibnamefont
  {Lynsky}}, \bibinfo {author} {\bibfnamefont {A.~I.}\ \bibnamefont
  {Alhassan}}, \bibinfo {author} {\bibfnamefont {G.}~\bibnamefont {Lheureux}},
  \bibinfo {author} {\bibfnamefont {B.}~\bibnamefont {Bonef}}, \bibinfo
  {author} {\bibfnamefont {S.~P.}\ \bibnamefont {DenBaars}}, \bibinfo {author}
  {\bibfnamefont {S.}~\bibnamefont {Nakamura}}, \bibinfo {author}
  {\bibfnamefont {Y.-R.}\ \bibnamefont {Wu}}, \bibinfo {author} {\bibfnamefont
  {C.}~\bibnamefont {Weisbuch}},\ and\ \bibinfo {author} {\bibfnamefont
  {J.~S.}\ \bibnamefont {Speck}},\ }\bibfield  {title} {\enquote {\bibinfo
  {title} {{Barriers to carrier transport in multiple quantum well
  nitride-based c-plane green light emitting diodes}},}\ }\href@noop {}
  {\bibfield  {journal} {\bibinfo  {journal} {Physical Review Materials}\
  }\textbf {\bibinfo {volume} {4}},\ \bibinfo {pages} {054604} (\bibinfo {year}
  {2020})}\BibitemShut {NoStop}%
\bibitem [{\citenamefont {Mukund}(2014)}]{mukund2014ingan}%
  \BibitemOpen
  \bibfield  {author} {\bibinfo {author} {\bibfnamefont {A.~H.}\ \bibnamefont
  {Mukund}},\ }\href@noop {} {\emph {\bibinfo {title} {{InGaN/GaN Multiple
  Quantum Well Light-Emitting Diodes grown on Polar, Semi-polar and Non-Polar
  Orientations}}}}\ (\bibinfo  {publisher} {North Carolina State University},\
  \bibinfo {year} {2014})\BibitemShut {NoStop}%
\bibitem [{\citenamefont {Lheureux}\ \emph {et~al.}(2020)\citenamefont
  {Lheureux}, \citenamefont {Lynsky}, \citenamefont {Wu}, \citenamefont
  {Speck},\ and\ \citenamefont {Weisbuch}}]{lheureux20203d}%
  \BibitemOpen
  \bibfield  {author} {\bibinfo {author} {\bibfnamefont {G.}~\bibnamefont
  {Lheureux}}, \bibinfo {author} {\bibfnamefont {C.}~\bibnamefont {Lynsky}},
  \bibinfo {author} {\bibfnamefont {Y.-R.}\ \bibnamefont {Wu}}, \bibinfo
  {author} {\bibfnamefont {J.~S.}\ \bibnamefont {Speck}},\ and\ \bibinfo
  {author} {\bibfnamefont {C.}~\bibnamefont {Weisbuch}},\ }\bibfield  {title}
  {\enquote {\bibinfo {title} {{A 3D simulation comparison of carrier transport
  in green and blue c-plane multi-quantum well nitride light emitting
  diodes}},}\ }\href@noop {} {\bibfield  {journal} {\bibinfo  {journal}
  {Journal of Applied Physics}\ }\textbf {\bibinfo {volume} {128}},\ \bibinfo
  {pages} {235703} (\bibinfo {year} {2020})}\BibitemShut {NoStop}%
\bibitem [{\citenamefont {Qwah}\ \emph {et~al.}(2020)\citenamefont {Qwah},
  \citenamefont {Monavarian}, \citenamefont {Lheureux}, \citenamefont {Wang},
  \citenamefont {Wu},\ and\ \citenamefont {Speck}}]{qwah2020theoretical}%
  \BibitemOpen
  \bibfield  {author} {\bibinfo {author} {\bibfnamefont {K.}~\bibnamefont
  {Qwah}}, \bibinfo {author} {\bibfnamefont {M.}~\bibnamefont {Monavarian}},
  \bibinfo {author} {\bibfnamefont {G.}~\bibnamefont {Lheureux}}, \bibinfo
  {author} {\bibfnamefont {J.}~\bibnamefont {Wang}}, \bibinfo {author}
  {\bibfnamefont {Y.-R.}\ \bibnamefont {Wu}},\ and\ \bibinfo {author}
  {\bibfnamefont {J.}~\bibnamefont {Speck}},\ }\bibfield  {title} {\enquote
  {\bibinfo {title} {{Theoretical and experimental investigations of vertical
  hole transport through unipolar AlGaN structures: impacts of random alloy
  disorder}},}\ }\href@noop {} {\bibfield  {journal} {\bibinfo  {journal}
  {Applied Physics Letters}\ }\textbf {\bibinfo {volume} {117}},\ \bibinfo
  {pages} {022107} (\bibinfo {year} {2020})}\BibitemShut {NoStop}%
\bibitem [{\citenamefont {Yang}\ \emph {et~al.}(2014)\citenamefont {Yang},
  \citenamefont {Shivaraman}, \citenamefont {Speck},\ and\ \citenamefont
  {Wu}}]{yang2014influence}%
  \BibitemOpen
  \bibfield  {author} {\bibinfo {author} {\bibfnamefont {T.-J.}\ \bibnamefont
  {Yang}}, \bibinfo {author} {\bibfnamefont {R.}~\bibnamefont {Shivaraman}},
  \bibinfo {author} {\bibfnamefont {J.~S.}\ \bibnamefont {Speck}},\ and\
  \bibinfo {author} {\bibfnamefont {Y.-R.}\ \bibnamefont {Wu}},\ }\bibfield
  {title} {\enquote {\bibinfo {title} {{The influence of random indium alloy
  fluctuations in indium gallium nitride quantum wells on the device
  behavior}},}\ }\href@noop {} {\bibfield  {journal} {\bibinfo  {journal}
  {Journal of Applied Physics}\ }\textbf {\bibinfo {volume} {116}},\ \bibinfo
  {pages} {113104} (\bibinfo {year} {2014})}\BibitemShut {NoStop}%
\bibitem [{\citenamefont {Wu}\ \emph {et~al.}(2012)\citenamefont {Wu},
  \citenamefont {Shivaraman}, \citenamefont {Wang},\ and\ \citenamefont
  {Speck}}]{wu2012analyzing}%
  \BibitemOpen
  \bibfield  {author} {\bibinfo {author} {\bibfnamefont {Y.-R.}\ \bibnamefont
  {Wu}}, \bibinfo {author} {\bibfnamefont {R.}~\bibnamefont {Shivaraman}},
  \bibinfo {author} {\bibfnamefont {K.-C.}\ \bibnamefont {Wang}},\ and\
  \bibinfo {author} {\bibfnamefont {J.~S.}\ \bibnamefont {Speck}},\ }\bibfield
  {title} {\enquote {\bibinfo {title} {{Analyzing the physical properties of
  InGaN multiple quantum well light emitting diodes from nano scale
  structure}},}\ }\href@noop {} {\bibfield  {journal} {\bibinfo  {journal}
  {Applied Physics Letters}\ }\textbf {\bibinfo {volume} {101}},\ \bibinfo
  {pages} {083505} (\bibinfo {year} {2012})}\BibitemShut {NoStop}%
\bibitem [{\citenamefont {Chen}\ \emph {et~al.}(2018)\citenamefont {Chen},
  \citenamefont {Speck}, \citenamefont {Weisbuch},\ and\ \citenamefont
  {Wu}}]{chen2018three}%
  \BibitemOpen
  \bibfield  {author} {\bibinfo {author} {\bibfnamefont {H.-H.}\ \bibnamefont
  {Chen}}, \bibinfo {author} {\bibfnamefont {J.~S.}\ \bibnamefont {Speck}},
  \bibinfo {author} {\bibfnamefont {C.}~\bibnamefont {Weisbuch}},\ and\
  \bibinfo {author} {\bibfnamefont {Y.-R.}\ \bibnamefont {Wu}},\ }\bibfield
  {title} {\enquote {\bibinfo {title} {{Three dimensional simulation on the
  transport and quantum efficiency of {UVC-LEDs} with random alloy
  fluctuations}},}\ }\href@noop {} {\bibfield  {journal} {\bibinfo  {journal}
  {Applied Physics Letters}\ }\textbf {\bibinfo {volume} {113}},\ \bibinfo
  {pages} {153504} (\bibinfo {year} {2018})}\BibitemShut {NoStop}%
\bibitem [{\citenamefont {der Maur}, \citenamefont {Pecchia},\ and\
  \citenamefont {Di~Carlo}(2015)}]{der2015influence}%
  \BibitemOpen
  \bibfield  {author} {\bibinfo {author} {\bibfnamefont {M.~A.}\ \bibnamefont
  {der Maur}}, \bibinfo {author} {\bibfnamefont {A.}~\bibnamefont {Pecchia}},\
  and\ \bibinfo {author} {\bibfnamefont {A.}~\bibnamefont {Di~Carlo}},\
  }\bibfield  {title} {\enquote {\bibinfo {title} {Influence of random alloy
  fluctuations in ingan/gan quantum wells on led efficiency},}\ }in\ \href@noop
  {} {\emph {\bibinfo {booktitle} {2015 IEEE 1st International Forum on
  Research and Technologies for Society and Industry Leveraging a better
  tomorrow (RTSI)}}}\ (\bibinfo {organization} {IEEE},\ \bibinfo {year}
  {2015})\ pp.\ \bibinfo {pages} {153--156}\BibitemShut {NoStop}%
\bibitem [{\citenamefont {Lester}\ \emph {et~al.}(1995)\citenamefont {Lester},
  \citenamefont {Ponce}, \citenamefont {Craford},\ and\ \citenamefont
  {Steigerwald}}]{lester1995high}%
  \BibitemOpen
  \bibfield  {author} {\bibinfo {author} {\bibfnamefont {S.}~\bibnamefont
  {Lester}}, \bibinfo {author} {\bibfnamefont {F.~A.}\ \bibnamefont {Ponce}},
  \bibinfo {author} {\bibfnamefont {M.~G.}\ \bibnamefont {Craford}},\ and\
  \bibinfo {author} {\bibfnamefont {D.~A.}\ \bibnamefont {Steigerwald}},\
  }\bibfield  {title} {\enquote {\bibinfo {title} {{High dislocation densities
  in high efficiency GaN-based light-emitting diodes}},}\ }\href@noop {}
  {\bibfield  {journal} {\bibinfo  {journal} {Applied Physics Letters}\
  }\textbf {\bibinfo {volume} {66}},\ \bibinfo {pages} {1249--1251} (\bibinfo
  {year} {1995})}\BibitemShut {NoStop}%
\bibitem [{\citenamefont {Voronenkov}\ \emph {et~al.}(2013)\citenamefont
  {Voronenkov}, \citenamefont {Bochkareva}, \citenamefont {Gorbunov},
  \citenamefont {Latyshev}, \citenamefont {Lelikov}, \citenamefont {Rebane},
  \citenamefont {Tsyuk}, \citenamefont {Zubrilov},\ and\ \citenamefont
  {Shreter}}]{voronenkov2013nature}%
  \BibitemOpen
  \bibfield  {author} {\bibinfo {author} {\bibfnamefont {V.}~\bibnamefont
  {Voronenkov}}, \bibinfo {author} {\bibfnamefont {N.}~\bibnamefont
  {Bochkareva}}, \bibinfo {author} {\bibfnamefont {R.}~\bibnamefont
  {Gorbunov}}, \bibinfo {author} {\bibfnamefont {P.}~\bibnamefont {Latyshev}},
  \bibinfo {author} {\bibfnamefont {Y.}~\bibnamefont {Lelikov}}, \bibinfo
  {author} {\bibfnamefont {Y.}~\bibnamefont {Rebane}}, \bibinfo {author}
  {\bibfnamefont {A.}~\bibnamefont {Tsyuk}}, \bibinfo {author} {\bibfnamefont
  {A.}~\bibnamefont {Zubrilov}},\ and\ \bibinfo {author} {\bibfnamefont
  {Y.}~\bibnamefont {Shreter}},\ }\bibfield  {title} {\enquote {\bibinfo
  {title} {{Nature of V-shaped defects in GaN}},}\ }\href@noop {} {\bibfield
  {journal} {\bibinfo  {journal} {Japanese Journal of Applied Physics}\
  }\textbf {\bibinfo {volume} {52}},\ \bibinfo {pages} {08JE14} (\bibinfo
  {year} {2013})}\BibitemShut {NoStop}%
\bibitem [{\citenamefont {Yoshida}\ \emph {et~al.}(2015)\citenamefont
  {Yoshida}, \citenamefont {Hikosaka}, \citenamefont {Nago},\ and\
  \citenamefont {Nunoue}}]{yoshida2015impact}%
  \BibitemOpen
  \bibfield  {author} {\bibinfo {author} {\bibfnamefont {H.}~\bibnamefont
  {Yoshida}}, \bibinfo {author} {\bibfnamefont {T.}~\bibnamefont {Hikosaka}},
  \bibinfo {author} {\bibfnamefont {H.}~\bibnamefont {Nago}},\ and\ \bibinfo
  {author} {\bibfnamefont {S.}~\bibnamefont {Nunoue}},\ }\bibfield  {title}
  {\enquote {\bibinfo {title} {{Impact of dislocations and defects on the
  relaxation behavior of InGaN/GaN multiple quantum wells grown on Si and
  sapphire substrates}},}\ }\href@noop {} {\bibfield  {journal} {\bibinfo
  {journal} {physica status solidi (b)}\ }\textbf {\bibinfo {volume} {252}},\
  \bibinfo {pages} {917--922} (\bibinfo {year} {2015})}\BibitemShut {NoStop}%
\bibitem [{\citenamefont {Li}\ \emph {et~al.}(2016)\citenamefont {Li},
  \citenamefont {Wu}, \citenamefont {Hsu}, \citenamefont {Lu}, \citenamefont
  {Li}, \citenamefont {Lu},\ and\ \citenamefont {Wu}}]{li20163d}%
  \BibitemOpen
  \bibfield  {author} {\bibinfo {author} {\bibfnamefont {C.-K.}\ \bibnamefont
  {Li}}, \bibinfo {author} {\bibfnamefont {C.-K.}\ \bibnamefont {Wu}}, \bibinfo
  {author} {\bibfnamefont {C.-C.}\ \bibnamefont {Hsu}}, \bibinfo {author}
  {\bibfnamefont {L.-S.}\ \bibnamefont {Lu}}, \bibinfo {author} {\bibfnamefont
  {H.}~\bibnamefont {Li}}, \bibinfo {author} {\bibfnamefont {T.-C.}\
  \bibnamefont {Lu}},\ and\ \bibinfo {author} {\bibfnamefont {Y.-R.}\
  \bibnamefont {Wu}},\ }\bibfield  {title} {\enquote {\bibinfo {title} {3d
  numerical modeling of the carrier transport and radiative efficiency for
  ingan/gan light emitting diodes with v-shaped pits},}\ }\href@noop {}
  {\bibfield  {journal} {\bibinfo  {journal} {AIP Advances}\ }\textbf {\bibinfo
  {volume} {6}},\ \bibinfo {pages} {055208} (\bibinfo {year}
  {2016})}\BibitemShut {NoStop}%
\bibitem [{\citenamefont {Cho}\ \emph {et~al.}(2013)\citenamefont {Cho},
  \citenamefont {Kim}, \citenamefont {Kim}, \citenamefont {Shim}, \citenamefont
  {Hwang}, \citenamefont {Park}, \citenamefont {Park},\ and\ \citenamefont
  {Kim}}]{cho2013quantum}%
  \BibitemOpen
  \bibfield  {author} {\bibinfo {author} {\bibfnamefont {Y.-H.}\ \bibnamefont
  {Cho}}, \bibinfo {author} {\bibfnamefont {J.-Y.}\ \bibnamefont {Kim}},
  \bibinfo {author} {\bibfnamefont {J.}~\bibnamefont {Kim}}, \bibinfo {author}
  {\bibfnamefont {M.-B.}\ \bibnamefont {Shim}}, \bibinfo {author}
  {\bibfnamefont {S.}~\bibnamefont {Hwang}}, \bibinfo {author} {\bibfnamefont
  {S.-H.}\ \bibnamefont {Park}}, \bibinfo {author} {\bibfnamefont {Y.-S.}\
  \bibnamefont {Park}},\ and\ \bibinfo {author} {\bibfnamefont
  {S.}~\bibnamefont {Kim}},\ }\bibfield  {title} {\enquote {\bibinfo {title}
  {{Quantum efficiency affected by localized carrier distribution near the
  V-defect in GaN based quantum well}},}\ }\href@noop {} {\bibfield  {journal}
  {\bibinfo  {journal} {Applied Physics Letters}\ }\textbf {\bibinfo {volume}
  {103}},\ \bibinfo {pages} {261101} (\bibinfo {year} {2013})}\BibitemShut
  {NoStop}%
\bibitem [{\citenamefont {Quan}\ \emph {et~al.}(2016)\citenamefont {Quan},
  \citenamefont {Liu}, \citenamefont {Fang}, \citenamefont {Wang},\ and\
  \citenamefont {Jiang}}]{quan2016effect}%
  \BibitemOpen
  \bibfield  {author} {\bibinfo {author} {\bibfnamefont {Z.}~\bibnamefont
  {Quan}}, \bibinfo {author} {\bibfnamefont {J.}~\bibnamefont {Liu}}, \bibinfo
  {author} {\bibfnamefont {F.}~\bibnamefont {Fang}}, \bibinfo {author}
  {\bibfnamefont {G.}~\bibnamefont {Wang}},\ and\ \bibinfo {author}
  {\bibfnamefont {F.}~\bibnamefont {Jiang}},\ }\bibfield  {title} {\enquote
  {\bibinfo {title} {{Effect of V-shaped Pit area ratio on quantum efficiency
  of blue InGaN/GaN multiple-quantum well light-emitting diodes}},}\
  }\href@noop {} {\bibfield  {journal} {\bibinfo  {journal} {Optical and
  Quantum Electronics}\ }\textbf {\bibinfo {volume} {48}},\ \bibinfo {pages}
  {195} (\bibinfo {year} {2016})}\BibitemShut {NoStop}%
\bibitem [{\citenamefont {Shen}\ \emph {et~al.}(2021)\citenamefont {Shen},
  \citenamefont {Weisbuch}, \citenamefont {Speck},\ and\ \citenamefont
  {Wu}}]{shen2021three}%
  \BibitemOpen
  \bibfield  {author} {\bibinfo {author} {\bibfnamefont {H.-T.}\ \bibnamefont
  {Shen}}, \bibinfo {author} {\bibfnamefont {C.}~\bibnamefont {Weisbuch}},
  \bibinfo {author} {\bibfnamefont {J.~S.}\ \bibnamefont {Speck}},\ and\
  \bibinfo {author} {\bibfnamefont {Y.-R.}\ \bibnamefont {Wu}},\ }\bibfield
  {title} {\enquote {\bibinfo {title} {{Three-Dimensional Modeling of
  Minority-Carrier Lateral Diffusion Length Including Random Alloy Fluctuations
  in (In, Ga) N and (Al, Ga) N Single Quantum Wells}},}\ }\href@noop {}
  {\bibfield  {journal} {\bibinfo  {journal} {Physical Review Applied}\
  }\textbf {\bibinfo {volume} {16}},\ \bibinfo {pages} {024054} (\bibinfo
  {year} {2021})}\BibitemShut {NoStop}%
\bibitem [{\citenamefont {Zhou}\ and\ \citenamefont
  {Liu}(2017)}]{zhou2017effect}%
  \BibitemOpen
  \bibfield  {author} {\bibinfo {author} {\bibfnamefont {S.}~\bibnamefont
  {Zhou}}\ and\ \bibinfo {author} {\bibfnamefont {X.}~\bibnamefont {Liu}},\
  }\bibfield  {title} {\enquote {\bibinfo {title} {{Effect of V-pits embedded
  InGaN/GaN superlattices on optical and electrical properties of GaN-based
  green light-emitting diodes}},}\ }\href@noop {} {\bibfield  {journal}
  {\bibinfo  {journal} {physica status solidi (a)}\ }\textbf {\bibinfo {volume}
  {214}},\ \bibinfo {pages} {1600782} (\bibinfo {year} {2017})}\BibitemShut
  {NoStop}%
\bibitem [{\citenamefont {Wu}\ \emph {et~al.}(1998)\citenamefont {Wu},
  \citenamefont {Elsass}, \citenamefont {Abare}, \citenamefont {Mack},
  \citenamefont {Keller}, \citenamefont {Petroff}, \citenamefont {DenBaars},
  \citenamefont {Speck},\ and\ \citenamefont {Rosner}}]{wu1998structural}%
  \BibitemOpen
  \bibfield  {author} {\bibinfo {author} {\bibfnamefont {X.}~\bibnamefont
  {Wu}}, \bibinfo {author} {\bibfnamefont {C.}~\bibnamefont {Elsass}}, \bibinfo
  {author} {\bibfnamefont {A.}~\bibnamefont {Abare}}, \bibinfo {author}
  {\bibfnamefont {M.}~\bibnamefont {Mack}}, \bibinfo {author} {\bibfnamefont
  {S.}~\bibnamefont {Keller}}, \bibinfo {author} {\bibfnamefont
  {P.}~\bibnamefont {Petroff}}, \bibinfo {author} {\bibfnamefont
  {S.}~\bibnamefont {DenBaars}}, \bibinfo {author} {\bibfnamefont
  {J.}~\bibnamefont {Speck}},\ and\ \bibinfo {author} {\bibfnamefont
  {S.}~\bibnamefont {Rosner}},\ }\bibfield  {title} {\enquote {\bibinfo {title}
  {{Structural origin of V-defects and correlation with localized excitonic
  centers in InGaN/GaN multiple quantum wells}},}\ }\href@noop {} {\bibfield
  {journal} {\bibinfo  {journal} {Applied Physics Letters}\ }\textbf {\bibinfo
  {volume} {72}},\ \bibinfo {pages} {692--694} (\bibinfo {year}
  {1998})}\BibitemShut {NoStop}%
\bibitem [{\citenamefont {Hu}\ \emph {et~al.}(2012)\citenamefont {Hu},
  \citenamefont {Farrell}, \citenamefont {Neufeld}, \citenamefont {Iza},
  \citenamefont {Cruz}, \citenamefont {Pfaff}, \citenamefont {Simeonov},
  \citenamefont {Keller}, \citenamefont {Nakamura}, \citenamefont {DenBaars},
  \citenamefont {Mishra},\ and\ \citenamefont {Speck}}]{hu2012effect}%
  \BibitemOpen
  \bibfield  {author} {\bibinfo {author} {\bibfnamefont {Y.-L.}\ \bibnamefont
  {Hu}}, \bibinfo {author} {\bibfnamefont {R.~M.}\ \bibnamefont {Farrell}},
  \bibinfo {author} {\bibfnamefont {C.~J.}\ \bibnamefont {Neufeld}}, \bibinfo
  {author} {\bibfnamefont {M.}~\bibnamefont {Iza}}, \bibinfo {author}
  {\bibfnamefont {S.~C.}\ \bibnamefont {Cruz}}, \bibinfo {author}
  {\bibfnamefont {N.}~\bibnamefont {Pfaff}}, \bibinfo {author} {\bibfnamefont
  {D.}~\bibnamefont {Simeonov}}, \bibinfo {author} {\bibfnamefont
  {S.}~\bibnamefont {Keller}}, \bibinfo {author} {\bibfnamefont
  {S.}~\bibnamefont {Nakamura}}, \bibinfo {author} {\bibfnamefont {S.~P.}\
  \bibnamefont {DenBaars}}, \bibinfo {author} {\bibfnamefont {U.~K.}\
  \bibnamefont {Mishra}},\ and\ \bibinfo {author} {\bibfnamefont {J.~S.}\
  \bibnamefont {Speck}},\ }\bibfield  {title} {\enquote {\bibinfo {title}
  {{Effect of quantum well cap layer thickness on the microstructure and
  performance of InGaN/GaN solar cells}},}\ }\href@noop {} {\bibfield
  {journal} {\bibinfo  {journal} {Applied Physics Letters}\ }\textbf {\bibinfo
  {volume} {100}},\ \bibinfo {pages} {161101} (\bibinfo {year}
  {2012})}\BibitemShut {NoStop}%
\bibitem [{\citenamefont {Romanov}\ \emph {et~al.}(2006)\citenamefont
  {Romanov}, \citenamefont {Baker}, \citenamefont {Nakamura}, \citenamefont
  {Speck},\ and\ \citenamefont {Group}}]{romanov2006strain}%
  \BibitemOpen
  \bibfield  {author} {\bibinfo {author} {\bibfnamefont {A.}~\bibnamefont
  {Romanov}}, \bibinfo {author} {\bibfnamefont {T.}~\bibnamefont {Baker}},
  \bibinfo {author} {\bibfnamefont {S.}~\bibnamefont {Nakamura}}, \bibinfo
  {author} {\bibfnamefont {J.}~\bibnamefont {Speck}},\ and\ \bibinfo {author}
  {\bibfnamefont {E.~U.}\ \bibnamefont {Group}},\ }\bibfield  {title} {\enquote
  {\bibinfo {title} {{Strain-induced polarization in wurtzite III-nitride
  semipolar layers}},}\ }\href@noop {} {\bibfield  {journal} {\bibinfo
  {journal} {Journal of Applied Physics}\ }\textbf {\bibinfo {volume} {100}},\
  \bibinfo {pages} {023522} (\bibinfo {year} {2006})}\BibitemShut {NoStop}%
\bibitem [{\citenamefont {Jiang}\ \emph {et~al.}(2019)\citenamefont {Jiang},
  \citenamefont {Zhang}, \citenamefont {Xu}, \citenamefont {Ding},
  \citenamefont {Wang}, \citenamefont {Wu}, \citenamefont {Wang}, \citenamefont
  {Mo}, \citenamefont {Quan}, \citenamefont {Guo}, \citenamefont {Zheng},
  \citenamefont {Pan},\ and\ \citenamefont {Liu}}]{jiang2019efficient}%
  \BibitemOpen
  \bibfield  {author} {\bibinfo {author} {\bibfnamefont {F.}~\bibnamefont
  {Jiang}}, \bibinfo {author} {\bibfnamefont {J.}~\bibnamefont {Zhang}},
  \bibinfo {author} {\bibfnamefont {L.}~\bibnamefont {Xu}}, \bibinfo {author}
  {\bibfnamefont {J.}~\bibnamefont {Ding}}, \bibinfo {author} {\bibfnamefont
  {G.}~\bibnamefont {Wang}}, \bibinfo {author} {\bibfnamefont {X.}~\bibnamefont
  {Wu}}, \bibinfo {author} {\bibfnamefont {X.}~\bibnamefont {Wang}}, \bibinfo
  {author} {\bibfnamefont {C.}~\bibnamefont {Mo}}, \bibinfo {author}
  {\bibfnamefont {Z.}~\bibnamefont {Quan}}, \bibinfo {author} {\bibfnamefont
  {X.}~\bibnamefont {Guo}}, \bibinfo {author} {\bibfnamefont {C.}~\bibnamefont
  {Zheng}}, \bibinfo {author} {\bibfnamefont {S.}~\bibnamefont {Pan}},\ and\
  \bibinfo {author} {\bibfnamefont {J.}~\bibnamefont {Liu}},\ }\bibfield
  {title} {\enquote {\bibinfo {title} {{Efficient InGaN-based
  yellow-light-emitting diodes}},}\ }\href@noop {} {\bibfield  {journal}
  {\bibinfo  {journal} {Photonics Research}\ }\textbf {\bibinfo {volume} {7}},\
  \bibinfo {pages} {144--148} (\bibinfo {year} {2019})}\BibitemShut {NoStop}%
\bibitem [{\citenamefont {Lynsky}\ \emph {et~al.}(2021)\citenamefont {Lynsky},
  \citenamefont {White}, \citenamefont {Chow}, \citenamefont {Ho},
  \citenamefont {Nakamura}, \citenamefont {DenBaars},\ and\ \citenamefont
  {Speck}}]{lynsky2021role}%
  \BibitemOpen
  \bibfield  {author} {\bibinfo {author} {\bibfnamefont {C.}~\bibnamefont
  {Lynsky}}, \bibinfo {author} {\bibfnamefont {R.~C.}\ \bibnamefont {White}},
  \bibinfo {author} {\bibfnamefont {Y.~C.}\ \bibnamefont {Chow}}, \bibinfo
  {author} {\bibfnamefont {W.~Y.}\ \bibnamefont {Ho}}, \bibinfo {author}
  {\bibfnamefont {S.}~\bibnamefont {Nakamura}}, \bibinfo {author}
  {\bibfnamefont {S.~P.}\ \bibnamefont {DenBaars}},\ and\ \bibinfo {author}
  {\bibfnamefont {J.~S.}\ \bibnamefont {Speck}},\ }\bibfield  {title} {\enquote
  {\bibinfo {title} {{Role of V-defect density on the performance of
  III-nitride green LEDs on sapphire substrates}},}\ }\href@noop {} {\bibfield
  {journal} {\bibinfo  {journal} {Journal of Crystal Growth}\ }\textbf
  {\bibinfo {volume} {560}},\ \bibinfo {pages} {126048} (\bibinfo {year}
  {2021})}\BibitemShut {NoStop}%
\bibitem [{\citenamefont {Han}\ \emph {et~al.}(2013)\citenamefont {Han},
  \citenamefont {Lee}, \citenamefont {Shim}, \citenamefont {Wook~Lee},
  \citenamefont {Kim}, \citenamefont {Yoon}, \citenamefont {Sun~Kim},\ and\
  \citenamefont {Kim}}]{han2013improvement}%
  \BibitemOpen
  \bibfield  {author} {\bibinfo {author} {\bibfnamefont {S.-H.}\ \bibnamefont
  {Han}}, \bibinfo {author} {\bibfnamefont {D.-Y.}\ \bibnamefont {Lee}},
  \bibinfo {author} {\bibfnamefont {H.-W.}\ \bibnamefont {Shim}}, \bibinfo
  {author} {\bibfnamefont {J.}~\bibnamefont {Wook~Lee}}, \bibinfo {author}
  {\bibfnamefont {D.-J.}\ \bibnamefont {Kim}}, \bibinfo {author} {\bibfnamefont
  {S.}~\bibnamefont {Yoon}}, \bibinfo {author} {\bibfnamefont {Y.}~\bibnamefont
  {Sun~Kim}},\ and\ \bibinfo {author} {\bibfnamefont {S.-T.}\ \bibnamefont
  {Kim}},\ }\bibfield  {title} {\enquote {\bibinfo {title} {{Improvement of
  efficiency and electrical properties using intentionally formed V-shaped pits
  in InGaN/GaN multiple quantum well light-emitting diodes}},}\ }\href@noop {}
  {\bibfield  {journal} {\bibinfo  {journal} {Applied Physics Letters}\
  }\textbf {\bibinfo {volume} {102}},\ \bibinfo {pages} {251123} (\bibinfo
  {year} {2013})}\BibitemShut {NoStop}%
\bibitem [{\citenamefont {Ren}\ \emph {et~al.}(2016)\citenamefont {Ren},
  \citenamefont {Rouet-Leduc}, \citenamefont {Griffiths}, \citenamefont
  {Bohacek}, \citenamefont {Wallace}, \citenamefont {Edwards}, \citenamefont
  {Hopkins}, \citenamefont {Allsopp}, \citenamefont {Kappers}, \citenamefont
  {Martin},\ and\ \citenamefont {Oliver}}]{ren2016analysis}%
  \BibitemOpen
  \bibfield  {author} {\bibinfo {author} {\bibfnamefont {C.}~\bibnamefont
  {Ren}}, \bibinfo {author} {\bibfnamefont {B.}~\bibnamefont {Rouet-Leduc}},
  \bibinfo {author} {\bibfnamefont {J.}~\bibnamefont {Griffiths}}, \bibinfo
  {author} {\bibfnamefont {E.}~\bibnamefont {Bohacek}}, \bibinfo {author}
  {\bibfnamefont {M.}~\bibnamefont {Wallace}}, \bibinfo {author} {\bibfnamefont
  {P.}~\bibnamefont {Edwards}}, \bibinfo {author} {\bibfnamefont
  {M.}~\bibnamefont {Hopkins}}, \bibinfo {author} {\bibfnamefont
  {D.}~\bibnamefont {Allsopp}}, \bibinfo {author} {\bibfnamefont
  {M.}~\bibnamefont {Kappers}}, \bibinfo {author} {\bibfnamefont
  {R.}~\bibnamefont {Martin}},\ and\ \bibinfo {author} {\bibfnamefont
  {R.}~\bibnamefont {Oliver}},\ }\bibfield  {title} {\enquote {\bibinfo {title}
  {{Analysis of defect-related inhomogeneous electroluminescence in InGaN/GaN
  QW LEDs}},}\ }\href@noop {} {\bibfield  {journal} {\bibinfo  {journal}
  {Superlattices and Microstructures}\ }\textbf {\bibinfo {volume} {99}},\
  \bibinfo {pages} {118--124} (\bibinfo {year} {2016})}\BibitemShut {NoStop}%
\bibitem [{\citenamefont {Bouveyron}, \citenamefont {Mrad},\ and\ \citenamefont
  {Charles}(2019)}]{bouveyron2019v}%
  \BibitemOpen
  \bibfield  {author} {\bibinfo {author} {\bibfnamefont {R.}~\bibnamefont
  {Bouveyron}}, \bibinfo {author} {\bibfnamefont {M.}~\bibnamefont {Mrad}},\
  and\ \bibinfo {author} {\bibfnamefont {M.}~\bibnamefont {Charles}},\
  }\bibfield  {title} {\enquote {\bibinfo {title} {{V-pit pinning at the
  interface of high and low-temperature gallium nitride growth}},}\ }\href@noop
  {} {\bibfield  {journal} {\bibinfo  {journal} {Japanese Journal of Applied
  Physics}\ }\textbf {\bibinfo {volume} {58}},\ \bibinfo {pages} {SC1035}
  (\bibinfo {year} {2019})}\BibitemShut {NoStop}%
\bibitem [{\citenamefont {Di~Vito}\ \emph {et~al.}(2020)\citenamefont
  {Di~Vito}, \citenamefont {Pecchia}, \citenamefont {Di~Carlo},\ and\
  \citenamefont {Auf~der Maur}}]{di2020simulating}%
  \BibitemOpen
  \bibfield  {author} {\bibinfo {author} {\bibfnamefont {A.}~\bibnamefont
  {Di~Vito}}, \bibinfo {author} {\bibfnamefont {A.}~\bibnamefont {Pecchia}},
  \bibinfo {author} {\bibfnamefont {A.}~\bibnamefont {Di~Carlo}},\ and\
  \bibinfo {author} {\bibfnamefont {M.}~\bibnamefont {Auf~der Maur}},\
  }\bibfield  {title} {\enquote {\bibinfo {title} {{Simulating random alloy
  effects in III-nitride light emitting diodes}},}\ }\href@noop {} {\bibfield
  {journal} {\bibinfo  {journal} {Journal of Applied Physics}\ }\textbf
  {\bibinfo {volume} {128}},\ \bibinfo {pages} {041102} (\bibinfo {year}
  {2020})}\BibitemShut {NoStop}%
\bibitem [{\citenamefont {Filoche}\ \emph {et~al.}(2017)\citenamefont
  {Filoche}, \citenamefont {Piccardo}, \citenamefont {Wu}, \citenamefont {Li},
  \citenamefont {Weisbuch},\ and\ \citenamefont
  {Mayboroda}}]{filoche2017localization}%
  \BibitemOpen
  \bibfield  {author} {\bibinfo {author} {\bibfnamefont {M.}~\bibnamefont
  {Filoche}}, \bibinfo {author} {\bibfnamefont {M.}~\bibnamefont {Piccardo}},
  \bibinfo {author} {\bibfnamefont {Y.-R.}\ \bibnamefont {Wu}}, \bibinfo
  {author} {\bibfnamefont {C.-K.}\ \bibnamefont {Li}}, \bibinfo {author}
  {\bibfnamefont {C.}~\bibnamefont {Weisbuch}},\ and\ \bibinfo {author}
  {\bibfnamefont {S.}~\bibnamefont {Mayboroda}},\ }\bibfield  {title} {\enquote
  {\bibinfo {title} {{Localization landscape theory of disorder in
  semiconductors. I. Theory and modeling}},}\ }\href@noop {} {\bibfield
  {journal} {\bibinfo  {journal} {Physical Review B}\ }\textbf {\bibinfo
  {volume} {95}},\ \bibinfo {pages} {144204} (\bibinfo {year}
  {2017})}\BibitemShut {NoStop}%
\bibitem [{\citenamefont {Piccardo}\ \emph {et~al.}(2017)\citenamefont
  {Piccardo}, \citenamefont {Li}, \citenamefont {Wu}, \citenamefont {Speck},
  \citenamefont {Bonef}, \citenamefont {Farrell}, \citenamefont {Filoche},
  \citenamefont {Martinelli}, \citenamefont {Peretti},\ and\ \citenamefont
  {Weisbuch}}]{PhysRevB.95.144205}%
  \BibitemOpen
  \bibfield  {author} {\bibinfo {author} {\bibfnamefont {M.}~\bibnamefont
  {Piccardo}}, \bibinfo {author} {\bibfnamefont {C.-K.}\ \bibnamefont {Li}},
  \bibinfo {author} {\bibfnamefont {Y.-R.}\ \bibnamefont {Wu}}, \bibinfo
  {author} {\bibfnamefont {J.~S.}\ \bibnamefont {Speck}}, \bibinfo {author}
  {\bibfnamefont {B.}~\bibnamefont {Bonef}}, \bibinfo {author} {\bibfnamefont
  {R.~M.}\ \bibnamefont {Farrell}}, \bibinfo {author} {\bibfnamefont
  {M.}~\bibnamefont {Filoche}}, \bibinfo {author} {\bibfnamefont
  {L.}~\bibnamefont {Martinelli}}, \bibinfo {author} {\bibfnamefont
  {J.}~\bibnamefont {Peretti}},\ and\ \bibinfo {author} {\bibfnamefont
  {C.}~\bibnamefont {Weisbuch}},\ }\bibfield  {title} {\enquote {\bibinfo
  {title} {Localization landscape theory of disorder in semiconductors. ii.
  urbach tails of disordered quantum well layers},}\ }\href
  {https://doi.org/10.1103/PhysRevB.95.144205} {\bibfield  {journal} {\bibinfo
  {journal} {Phys. Rev. B}\ }\textbf {\bibinfo {volume} {95}},\ \bibinfo
  {pages} {144205} (\bibinfo {year} {2017})}\BibitemShut {NoStop}%
\bibitem [{\citenamefont {Li}\ \emph {et~al.}(2017)\citenamefont {Li},
  \citenamefont {Piccardo}, \citenamefont {Lu}, \citenamefont {Mayboroda},
  \citenamefont {Martinelli}, \citenamefont {Peretti}, \citenamefont {Speck},
  \citenamefont {Weisbuch}, \citenamefont {Filoche},\ and\ \citenamefont
  {Wu}}]{li2017localization}%
  \BibitemOpen
  \bibfield  {author} {\bibinfo {author} {\bibfnamefont {C.-K.}\ \bibnamefont
  {Li}}, \bibinfo {author} {\bibfnamefont {M.}~\bibnamefont {Piccardo}},
  \bibinfo {author} {\bibfnamefont {L.-S.}\ \bibnamefont {Lu}}, \bibinfo
  {author} {\bibfnamefont {S.}~\bibnamefont {Mayboroda}}, \bibinfo {author}
  {\bibfnamefont {L.}~\bibnamefont {Martinelli}}, \bibinfo {author}
  {\bibfnamefont {J.}~\bibnamefont {Peretti}}, \bibinfo {author} {\bibfnamefont
  {J.~S.}\ \bibnamefont {Speck}}, \bibinfo {author} {\bibfnamefont
  {C.}~\bibnamefont {Weisbuch}}, \bibinfo {author} {\bibfnamefont
  {M.}~\bibnamefont {Filoche}},\ and\ \bibinfo {author} {\bibfnamefont {Y.-R.}\
  \bibnamefont {Wu}},\ }\bibfield  {title} {\enquote {\bibinfo {title}
  {{Localization landscape theory of disorder in semiconductors. III.
  Application to carrier transport and recombination in light emitting
  diodes}},}\ }\href@noop {} {\bibfield  {journal} {\bibinfo  {journal}
  {Physical Review B}\ }\textbf {\bibinfo {volume} {95}},\ \bibinfo {pages}
  {144206} (\bibinfo {year} {2017})}\BibitemShut {NoStop}%
\bibitem [{\citenamefont {Arnold}\ \emph {et~al.}(2016)\citenamefont {Arnold},
  \citenamefont {David}, \citenamefont {Jerison}, \citenamefont {Mayboroda},\
  and\ \citenamefont {Filoche}}]{PhysRevLett.116.056602}%
  \BibitemOpen
  \bibfield  {author} {\bibinfo {author} {\bibfnamefont {D.~N.}\ \bibnamefont
  {Arnold}}, \bibinfo {author} {\bibfnamefont {G.}~\bibnamefont {David}},
  \bibinfo {author} {\bibfnamefont {D.}~\bibnamefont {Jerison}}, \bibinfo
  {author} {\bibfnamefont {S.}~\bibnamefont {Mayboroda}},\ and\ \bibinfo
  {author} {\bibfnamefont {M.}~\bibnamefont {Filoche}},\ }\bibfield  {title}
  {\enquote {\bibinfo {title} {{Effective Confining Potential of Quantum States
  in Disordered Media}},}\ }\href
  {https://doi.org/10.1103/PhysRevLett.116.056602} {\bibfield  {journal}
  {\bibinfo  {journal} {Phys. Rev. Lett.}\ }\textbf {\bibinfo {volume} {116}},\
  \bibinfo {pages} {056602} (\bibinfo {year} {2016})}\BibitemShut {NoStop}%
\bibitem [{\citenamefont {Filoche}\ and\ \citenamefont
  {Mayboroda}(2012)}]{filoche2012universal}%
  \BibitemOpen
  \bibfield  {author} {\bibinfo {author} {\bibfnamefont {M.}~\bibnamefont
  {Filoche}}\ and\ \bibinfo {author} {\bibfnamefont {S.}~\bibnamefont
  {Mayboroda}},\ }\bibfield  {title} {\enquote {\bibinfo {title} {{Universal
  mechanism for Anderson and weak localization}},}\ }\href@noop {} {\bibfield
  {journal} {\bibinfo  {journal} {Proceedings of the National Academy of
  Sciences}\ }\textbf {\bibinfo {volume} {109}},\ \bibinfo {pages}
  {14761--14766} (\bibinfo {year} {2012})}\BibitemShut {NoStop}%
\bibitem [{\citenamefont {Kioupakis}\ \emph {et~al.}(2011)\citenamefont
  {Kioupakis}, \citenamefont {Rinke}, \citenamefont {Delaney},\ and\
  \citenamefont {Van~de Walle}}]{kioupakis2011indirect}%
  \BibitemOpen
  \bibfield  {author} {\bibinfo {author} {\bibfnamefont {E.}~\bibnamefont
  {Kioupakis}}, \bibinfo {author} {\bibfnamefont {P.}~\bibnamefont {Rinke}},
  \bibinfo {author} {\bibfnamefont {K.~T.}\ \bibnamefont {Delaney}},\ and\
  \bibinfo {author} {\bibfnamefont {C.~G.}\ \bibnamefont {Van~de Walle}},\
  }\bibfield  {title} {\enquote {\bibinfo {title} {{Indirect Auger
  recombination as a cause of efficiency droop in nitride light-emitting
  diodes}},}\ }\href@noop {} {\bibfield  {journal} {\bibinfo  {journal}
  {Applied Physics Letters}\ }\textbf {\bibinfo {volume} {98}},\ \bibinfo
  {pages} {161107} (\bibinfo {year} {2011})}\BibitemShut {NoStop}%
\bibitem [{\citenamefont {Le}\ \emph {et~al.}(2012)\citenamefont {Le},
  \citenamefont {Zhao}, \citenamefont {Jiang}, \citenamefont {Li},
  \citenamefont {Wu}, \citenamefont {Chen}, \citenamefont {Liu}, \citenamefont
  {Li}, \citenamefont {Fan}, \citenamefont {Zhu}, \citenamefont {Wang},
  \citenamefont {Zhang},\ and\ \citenamefont {Yang}}]{le2012carriers}%
  \BibitemOpen
  \bibfield  {author} {\bibinfo {author} {\bibfnamefont {L.~C.}\ \bibnamefont
  {Le}}, \bibinfo {author} {\bibfnamefont {D.~G.}\ \bibnamefont {Zhao}},
  \bibinfo {author} {\bibfnamefont {D.~S.}\ \bibnamefont {Jiang}}, \bibinfo
  {author} {\bibfnamefont {L.}~\bibnamefont {Li}}, \bibinfo {author}
  {\bibfnamefont {L.~L.}\ \bibnamefont {Wu}}, \bibinfo {author} {\bibfnamefont
  {P.}~\bibnamefont {Chen}}, \bibinfo {author} {\bibfnamefont {Z.~S.}\
  \bibnamefont {Liu}}, \bibinfo {author} {\bibfnamefont {Z.~C.}\ \bibnamefont
  {Li}}, \bibinfo {author} {\bibfnamefont {Y.~M.}\ \bibnamefont {Fan}},
  \bibinfo {author} {\bibfnamefont {J.~J.}\ \bibnamefont {Zhu}}, \bibinfo
  {author} {\bibfnamefont {H.}~\bibnamefont {Wang}}, \bibinfo {author}
  {\bibfnamefont {S.~M.}\ \bibnamefont {Zhang}},\ and\ \bibinfo {author}
  {\bibfnamefont {H.}~\bibnamefont {Yang}},\ }\bibfield  {title} {\enquote
  {\bibinfo {title} {{Carriers capturing of V-defect and its effect on leakage
  current and electroluminescence in InGaN-based light-emitting diodes}},}\
  }\href@noop {} {\bibfield  {journal} {\bibinfo  {journal} {Applied Physics
  Letters}\ }\textbf {\bibinfo {volume} {101}},\ \bibinfo {pages} {252110}
  (\bibinfo {year} {2012})}\BibitemShut {NoStop}%
\bibitem [{\citenamefont {Wu}\ \emph {et~al.}(2014)\citenamefont {Wu},
  \citenamefont {Liu}, \citenamefont {Quan}, \citenamefont {Xiong},
  \citenamefont {Zheng}, \citenamefont {Zhang}, \citenamefont {Mao},\ and\
  \citenamefont {Jiang}}]{wu2014electroluminescence}%
  \BibitemOpen
  \bibfield  {author} {\bibinfo {author} {\bibfnamefont {X.}~\bibnamefont
  {Wu}}, \bibinfo {author} {\bibfnamefont {J.}~\bibnamefont {Liu}}, \bibinfo
  {author} {\bibfnamefont {Z.}~\bibnamefont {Quan}}, \bibinfo {author}
  {\bibfnamefont {C.}~\bibnamefont {Xiong}}, \bibinfo {author} {\bibfnamefont
  {C.}~\bibnamefont {Zheng}}, \bibinfo {author} {\bibfnamefont
  {J.}~\bibnamefont {Zhang}}, \bibinfo {author} {\bibfnamefont
  {Q.}~\bibnamefont {Mao}},\ and\ \bibinfo {author} {\bibfnamefont
  {F.}~\bibnamefont {Jiang}},\ }\bibfield  {title} {\enquote {\bibinfo {title}
  {{Electroluminescence from the sidewall quantum wells in the V-shaped pits of
  InGaN light emitting diodes}},}\ }\href@noop {} {\bibfield  {journal}
  {\bibinfo  {journal} {Applied Physics Letters}\ }\textbf {\bibinfo {volume}
  {104}},\ \bibinfo {pages} {221101} (\bibinfo {year} {2014})}\BibitemShut
  {NoStop}%
\bibitem [{\citenamefont {Shiojiri}\ \emph {et~al.}(2006)\citenamefont
  {Shiojiri}, \citenamefont {Chuo}, \citenamefont {Hsu}, \citenamefont {Yang},\
  and\ \citenamefont {Saijo}}]{shiojiri2006structure}%
  \BibitemOpen
  \bibfield  {author} {\bibinfo {author} {\bibfnamefont {M.}~\bibnamefont
  {Shiojiri}}, \bibinfo {author} {\bibfnamefont {C.}~\bibnamefont {Chuo}},
  \bibinfo {author} {\bibfnamefont {J.}~\bibnamefont {Hsu}}, \bibinfo {author}
  {\bibfnamefont {J.}~\bibnamefont {Yang}},\ and\ \bibinfo {author}
  {\bibfnamefont {H.}~\bibnamefont {Saijo}},\ }\bibfield  {title} {\enquote
  {\bibinfo {title} {{Structure and formation mechanism of V defects in
  multiple In Ga N/ Ga N quantum well layers}},}\ }\href@noop {} {\bibfield
  {journal} {\bibinfo  {journal} {Journal of applied physics}\ }\textbf
  {\bibinfo {volume} {99}},\ \bibinfo {pages} {073505} (\bibinfo {year}
  {2006})}\BibitemShut {NoStop}%
\bibitem [{\citenamefont {Chang}\ \emph {et~al.}(2015)\citenamefont {Chang},
  \citenamefont {Li}, \citenamefont {Shih},\ and\ \citenamefont
  {Lu}}]{chang2015manipulation}%
  \BibitemOpen
  \bibfield  {author} {\bibinfo {author} {\bibfnamefont {C.-Y.}\ \bibnamefont
  {Chang}}, \bibinfo {author} {\bibfnamefont {H.}~\bibnamefont {Li}}, \bibinfo
  {author} {\bibfnamefont {Y.-T.}\ \bibnamefont {Shih}},\ and\ \bibinfo
  {author} {\bibfnamefont {T.-C.}\ \bibnamefont {Lu}},\ }\bibfield  {title}
  {\enquote {\bibinfo {title} {{Manipulation of nanoscale V-pits to optimize
  internal quantum efficiency of InGaN multiple quantum wells}},}\ }\href@noop
  {} {\bibfield  {journal} {\bibinfo  {journal} {Applied Physics Letters}\
  }\textbf {\bibinfo {volume} {106}},\ \bibinfo {pages} {091104} (\bibinfo
  {year} {2015})}\BibitemShut {NoStop}%
\bibitem [{\citenamefont {Tomiya}\ \emph {et~al.}(2011)\citenamefont {Tomiya},
  \citenamefont {Kanitani}, \citenamefont {Tanaka}, \citenamefont {Ohkubo},\
  and\ \citenamefont {Hono}}]{tomiya2011atomic}%
  \BibitemOpen
  \bibfield  {author} {\bibinfo {author} {\bibfnamefont {S.}~\bibnamefont
  {Tomiya}}, \bibinfo {author} {\bibfnamefont {Y.}~\bibnamefont {Kanitani}},
  \bibinfo {author} {\bibfnamefont {S.}~\bibnamefont {Tanaka}}, \bibinfo
  {author} {\bibfnamefont {T.}~\bibnamefont {Ohkubo}},\ and\ \bibinfo {author}
  {\bibfnamefont {K.}~\bibnamefont {Hono}},\ }\bibfield  {title} {\enquote
  {\bibinfo {title} {{Atomic scale characterization of GaInN/GaN multiple
  quantum wells in V-shaped pits}},}\ }\href@noop {} {\bibfield  {journal}
  {\bibinfo  {journal} {Applied Physics Letters}\ }\textbf {\bibinfo {volume}
  {98}},\ \bibinfo {pages} {181904} (\bibinfo {year} {2011})}\BibitemShut
  {NoStop}%
\bibitem [{\citenamefont {Zhou}\ \emph {et~al.}(2018)\citenamefont {Zhou},
  \citenamefont {Liu}, \citenamefont {Yan}, \citenamefont {Gao}, \citenamefont
  {Xu}, \citenamefont {Zhao}, \citenamefont {Quan}, \citenamefont {Gui},\ and\
  \citenamefont {Liu}}]{zhou2018effect}%
  \BibitemOpen
  \bibfield  {author} {\bibinfo {author} {\bibfnamefont {S.}~\bibnamefont
  {Zhou}}, \bibinfo {author} {\bibfnamefont {X.}~\bibnamefont {Liu}}, \bibinfo
  {author} {\bibfnamefont {H.}~\bibnamefont {Yan}}, \bibinfo {author}
  {\bibfnamefont {Y.}~\bibnamefont {Gao}}, \bibinfo {author} {\bibfnamefont
  {H.}~\bibnamefont {Xu}}, \bibinfo {author} {\bibfnamefont {J.}~\bibnamefont
  {Zhao}}, \bibinfo {author} {\bibfnamefont {Z.}~\bibnamefont {Quan}}, \bibinfo
  {author} {\bibfnamefont {C.}~\bibnamefont {Gui}},\ and\ \bibinfo {author}
  {\bibfnamefont {S.}~\bibnamefont {Liu}},\ }\bibfield  {title} {\enquote
  {\bibinfo {title} {{The effect of nanometre-scale V-pits on electronic and
  optical properties and efficiency droop of GaN-based green light-emitting
  diodes}},}\ }\href@noop {} {\bibfield  {journal} {\bibinfo  {journal}
  {Scientific reports}\ }\textbf {\bibinfo {volume} {8}},\ \bibinfo {pages}
  {1--12} (\bibinfo {year} {2018})}\BibitemShut {NoStop}%
\bibitem [{\citenamefont {Kumakura}\ \emph {et~al.}(2005)\citenamefont
  {Kumakura}, \citenamefont {Makimoto}, \citenamefont {Kobayashi},
  \citenamefont {Hashizume}, \citenamefont {Fukui},\ and\ \citenamefont
  {Hasegawa}}]{kumakura2005minority}%
  \BibitemOpen
  \bibfield  {author} {\bibinfo {author} {\bibfnamefont {K.}~\bibnamefont
  {Kumakura}}, \bibinfo {author} {\bibfnamefont {T.}~\bibnamefont {Makimoto}},
  \bibinfo {author} {\bibfnamefont {N.}~\bibnamefont {Kobayashi}}, \bibinfo
  {author} {\bibfnamefont {T.}~\bibnamefont {Hashizume}}, \bibinfo {author}
  {\bibfnamefont {T.}~\bibnamefont {Fukui}},\ and\ \bibinfo {author}
  {\bibfnamefont {H.}~\bibnamefont {Hasegawa}},\ }\bibfield  {title} {\enquote
  {\bibinfo {title} {{Minority carrier diffusion length in GaN: Dislocation
  density and doping concentration dependence}},}\ }\href@noop {} {\bibfield
  {journal} {\bibinfo  {journal} {Applied Physics Letters}\ }\textbf {\bibinfo
  {volume} {86}},\ \bibinfo {pages} {052105} (\bibinfo {year}
  {2005})}\BibitemShut {NoStop}%
\bibitem [{\citenamefont {Karpov}\ and\ \citenamefont
  {Makarov}(2002)}]{karpov2002dislocation}%
  \BibitemOpen
  \bibfield  {author} {\bibinfo {author} {\bibfnamefont {S.~Y.}\ \bibnamefont
  {Karpov}}\ and\ \bibinfo {author} {\bibfnamefont {Y.~N.}\ \bibnamefont
  {Makarov}},\ }\bibfield  {title} {\enquote {\bibinfo {title} {{Dislocation
  effect on light emission efficiency in gallium nitride}},}\ }\href@noop {}
  {\bibfield  {journal} {\bibinfo  {journal} {Applied Physics Letters}\
  }\textbf {\bibinfo {volume} {81}},\ \bibinfo {pages} {4721--4723} (\bibinfo
  {year} {2002})}\BibitemShut {NoStop}%
\bibitem [{\citenamefont {Robertson}\ \emph {et~al.}(2019)\citenamefont
  {Robertson}, \citenamefont {Qwah}, \citenamefont {Wu},\ and\ \citenamefont
  {Speck}}]{Robertson_2019}%
  \BibitemOpen
  \bibfield  {author} {\bibinfo {author} {\bibfnamefont {C.~A.}\ \bibnamefont
  {Robertson}}, \bibinfo {author} {\bibfnamefont {K.~S.}\ \bibnamefont {Qwah}},
  \bibinfo {author} {\bibfnamefont {Y.-R.}\ \bibnamefont {Wu}},\ and\ \bibinfo
  {author} {\bibfnamefont {J.~S.}\ \bibnamefont {Speck}},\ }\bibfield  {title}
  {\enquote {\bibinfo {title} {Modeling dislocation-related leakage currents in
  {GaN} p-n diodes},}\ }\href {https://doi.org/10.1063/1.5123394} {\bibfield
  {journal} {\bibinfo  {journal} {Journal of Applied Physics}\ }\textbf
  {\bibinfo {volume} {126}},\ \bibinfo {pages} {245705} (\bibinfo {year}
  {2019})},\ \Eprint {https://arxiv.org/abs/https://doi.org/10.1063/1.5123394}
  {https://doi.org/10.1063/1.5123394} \BibitemShut {NoStop}%
\bibitem [{\citenamefont {Qwah}\ \emph {et~al.}(2021)\citenamefont {Qwah},
  \citenamefont {Robertson}, \citenamefont {Wu},\ and\ \citenamefont
  {Speck}}]{Qwah_2021}%
  \BibitemOpen
  \bibfield  {author} {\bibinfo {author} {\bibfnamefont {K.~S.}\ \bibnamefont
  {Qwah}}, \bibinfo {author} {\bibfnamefont {C.~A.}\ \bibnamefont {Robertson}},
  \bibinfo {author} {\bibfnamefont {Y.-R.}\ \bibnamefont {Wu}},\ and\ \bibinfo
  {author} {\bibfnamefont {J.}~\bibnamefont {Speck}},\ }\bibfield  {title}
  {\enquote {\bibinfo {title} {Modeling dislocation-related reverse bias
  leakage in {GaN} p-n diodes},}\ }\href
  {https://doi.org/10.1088/1361-6641/abfdfc} {\bibfield  {journal} {\bibinfo
  {journal} {Semiconductor Science and Technology}\ } (\bibinfo {year}
  {2021}),\ 10.1088/1361-6641/abfdfc}\BibitemShut {NoStop}%
\bibitem [{\citenamefont {Ryu}, \citenamefont {Kim},\ and\ \citenamefont
  {Shim}(2009)}]{ryu2009rate}%
  \BibitemOpen
  \bibfield  {author} {\bibinfo {author} {\bibfnamefont {H.-Y.}\ \bibnamefont
  {Ryu}}, \bibinfo {author} {\bibfnamefont {H.-S.}\ \bibnamefont {Kim}},\ and\
  \bibinfo {author} {\bibfnamefont {J.-I.}\ \bibnamefont {Shim}},\ }\bibfield
  {title} {\enquote {\bibinfo {title} {{Rate equation analysis of efficiency
  droop in InGaN light-emitting diodes}},}\ }\href@noop {} {\bibfield
  {journal} {\bibinfo  {journal} {Applied Physics Letters}\ }\textbf {\bibinfo
  {volume} {95}},\ \bibinfo {pages} {081114} (\bibinfo {year}
  {2009})}\BibitemShut {NoStop}%
\bibitem [{\citenamefont {Wang}, \citenamefont {Chiou},\ and\ \citenamefont
  {Chang}(2019)}]{wang2019investigating}%
  \BibitemOpen
  \bibfield  {author} {\bibinfo {author} {\bibfnamefont {C.}~\bibnamefont
  {Wang}}, \bibinfo {author} {\bibfnamefont {Y.}~\bibnamefont {Chiou}},\ and\
  \bibinfo {author} {\bibfnamefont {H.}~\bibnamefont {Chang}},\ }\bibfield
  {title} {\enquote {\bibinfo {title} {{Investigating the Efficiency Droop of
  Nitride-Based Blue LEDs with Different Quantum Barrier Growth Rates}},}\
  }\href@noop {} {\bibfield  {journal} {\bibinfo  {journal} {Crystals}\
  }\textbf {\bibinfo {volume} {9}},\ \bibinfo {pages} {677} (\bibinfo {year}
  {2019})}\BibitemShut {NoStop}%
\bibitem [{\citenamefont {Li}\ \emph {et~al.}(2019)\citenamefont {Li},
  \citenamefont {Tang}, \citenamefont {Zhang}, \citenamefont {Guo},
  \citenamefont {Li}, \citenamefont {Su}, \citenamefont {Li},\ and\
  \citenamefont {Yun}}]{li2019nanoscale}%
  \BibitemOpen
  \bibfield  {author} {\bibinfo {author} {\bibfnamefont {Y.}~\bibnamefont
  {Li}}, \bibinfo {author} {\bibfnamefont {W.}~\bibnamefont {Tang}}, \bibinfo
  {author} {\bibfnamefont {Y.}~\bibnamefont {Zhang}}, \bibinfo {author}
  {\bibfnamefont {M.}~\bibnamefont {Guo}}, \bibinfo {author} {\bibfnamefont
  {Q.}~\bibnamefont {Li}}, \bibinfo {author} {\bibfnamefont {X.}~\bibnamefont
  {Su}}, \bibinfo {author} {\bibfnamefont {A.}~\bibnamefont {Li}},\ and\
  \bibinfo {author} {\bibfnamefont {F.}~\bibnamefont {Yun}},\ }\bibfield
  {title} {\enquote {\bibinfo {title} {{Nanoscale characterization of V-defect
  in InGaN/GaN QWs LEDs using near-field scanning optical microscopy}},}\
  }\href@noop {} {\bibfield  {journal} {\bibinfo  {journal} {Nanomaterials}\
  }\textbf {\bibinfo {volume} {9}},\ \bibinfo {pages} {633} (\bibinfo {year}
  {2019})}\BibitemShut {NoStop}%
\end{thebibliography}%

\end{document}